\documentstyle[11pt,epsfig]{article}
\newcommand{\blankline}{\vskip .3cm}
\newcommand{\f}{\begin{equation}}
\newcommand{\ff}{\end{equation}}
\begin{document}
\centerline{\LARGE  A candidate for a background independent 
formulation of $\cal M$ theory}
\blankline
\centerline{Lee Smolin${}^*$}
\blankline
\centerline{\it  Center for Gravitational Physics and Geometry}
\centerline{\it Department of Physics}
\centerline {\it The Pennsylvania State University}
\centerline{\it University Park, PA, USA 16802}
\vfill
\centerline{March 10, 1999, revised Dec 26,1999}
\vfill
\centerline{ABSTRACT}
A class of background independent membrane field theories are studied,
and several properties are discovered which suggest
that they may play a role in 
a background independent form of $\cal M$ theory.  The bulk kinematics
of these theories are described in terms of the conformal blocks
of an algebra $G_{q}$ on all oriented, finite genus, two-surfaces.
The bulk dynamics is described in terms of causal histories in which
time evolution is specified by giving amplitudes to certain local
changes of the states. Holographic observables are defined
which live in finite dimensional states spaces associated with
boundaries in spacetime.  We show here that the natural observables
in these boundary state spaces are, when $G_{q}$ is chosen to be
$Spin(D)$ or a supersymmetric extension of it, 
generalizations of matrix model coordinates in $D$ dimensions.  In
certain cases the bulk dynamics can be chosen so the matrix model
dynamics is recoverd for the boundary observables. The bosonic and
supersymmetric cases in $D=3$ and $D=9$ are studied, and it is shown
that the latter is, in a certain limit, related to the matrix
model formulation of $\cal M$ theory.  This correspondence gives rise 
to a conjecture concerning a background independent form of $\cal M$ 
theory in terms of which excitations of the background independent membrane field 
theory that correspond to strings and $D0$ branes are identified.

\vfill
\blankline
${}^*$ smolin@phys.psu.edu
\eject

\section{Introduction}

In this paper a proposal is made for a background independent
formulation of $\cal M$ theory\cite{mtheory}.  This is based on a particular 
case of
a general formulation for a background
independent membrane field theory previously proposed with
Markopoulou\cite{tubes}.  That work was, in turn, a natural
extension of the spin network formalism that has been shown
to characterize the exact gauge
and diffeomorphism invariant states of a large class of theories
of quantum gravity\cite{lp1,lp2,sn1,sn2,spain,carlo-review},
including supergravity\cite{superstuff,yilee1}.

The spin network states are background independent, in the
sense that they make no reference to any background
metric or fields, because
they result from an exact, non-perturbative quantization
of the commutation relations of the gravitational fields. 
However, as they are constructed through a canonical
quantization procedure, they do 
depend on the background topological and differential structure.
As $S$ duality\cite{mtheory} and mirror manifolds\cite{mirror} 
in string theory indicates the presence of symmetries mixing
manifolds with different topology, this dependence must also be
eliminated if one wants to construct a successful background
independent form of $\cal M$ theory.  In \cite{tubes} it was
shown that this can be accomplished simply by thickening the
graphs which underlie the spin network states so that they become
membranes.  The labels on the spin networks are then replaced by 
the   
conformal
blocks ${\cal V}_{g}$ of a chiral theory based on a quantum
group or supergroup, $G_{q}$ for a two-manifold, ${\cal S}_{g}$ 
of genus $g$.  The full space of states is then the conformal blocks of
{\it all} compact oriented two-surfaces,
\f
{\cal H}=\sum_{g} {\cal V}_{g} .
\label{Hdef}
\ff
In quantum general 
relativity\cite{chopinlee,linking}, in $3+1$ the level is proportional
to the inverse cosmological constant,
\f
k= {6 \pi \over G^{2}\Lambda}
\label{cc}
\ff
As a consequence, the theory reduces to the spin network
formalism in the limit that the cosmological constant, $\Lambda$,
vanishes

Because the graphs of the spin network
formalism have been thickened to two-surfaces, no embedding manifold
need be assumed because there are states in $\cal H$  that carry appropriate
information to construct both a manifold (of any dimension and 
topology) and the embedding of a two surface in it\cite{tubes}.
In fact, as we shall discuss below, 
what is naturally constructed are pseudomanifolds, whose
defects may be the background independent analogues of branes\cite{stulee}.   

In \cite{pqtubes,towards} it was suggested
that because of its resemblance to a background independent
membrane theory, and the fact that such a theory may have a set of 
dynamically determined continuum
limits which are classical spacetimes of different dimension and
topology, an appropriately chosen member of this class of 
theories might 
be a background independent form of $\cal M$ theory.  In \cite{pqtubes} it
was also shown that a theory of this kind can incorporate
$SL(2,Z)$ string duality symmetry, leading to a non-perturbative
description of string network states.  
In this paper a new connection between the background independent 
membrane theories described in
\cite{tubes} and string theory is reported, which leads to a proposal
for a background independent form of $\cal M$ theory.

This new connection is the discovery of a certain class of observables,
which are related to the coordinates of matrix model descriptions of
membrane theory. 
As we will see, the bosonic matrix model in
$D$ dimensions can be recovered when $G_{q}$ is chosen to be the
quantum deformation of $Spin(D)$ and a certain choice of the dynamics
is made. In certain cases this can be extended to a supersymmetric
matrix model in $D$ dimensions, by exending $G_{q}$ to a superalgebra 
which has a  $Spin(D)$ subalgebra. In the $9$ dimensional case 
relevant for the 
deWit-Hoppe-Nicolai-Banks-Fischler-Shenker-Susskind \cite{dWHN,BFSS} matrix 
model the appropriate superalgebra is, as we will discuss, $SU(16|1)$.

The natural physical interpretation of these 
observables which is suggested here is that they are the background
independent versions of the coordinates of $D0$ branes. This is 
because we are able to argue that {\it if} the theory, in the form
given below, has a semiclassical limit which is flat $11$ dimensional
spacetime, those observables will indeed give the positions in the
$9$ dimensional transverse space, defined by the light cone 
coordinates of a physical observer, of zero-dimensional objects
on which strings end.  

To see how this description emerges, it is necessary to understand
the way in which the holographic principle is expressed in the
background indepenedent theories described in \cite{tubes}.
In fact, the holographic principle, in a cosmological form
discussed in \cite{tubes,hic,screens}, is a major 
part of the proposal made in
\cite{tubes}. As this theory has neither a background spacetime nor
asymptotic or external regions, it is naturally cosmological.
There is no place outside of the dynamical system for the
observer to live.  As was discussed in \cite{pluralism}, in such a case
all the observables of the quantum theory will be associated 
with closed surfaces that are dynamically embedded in the spacetime.  
These observables describe what an
observer living on the surface could measure about the physics
on its interior.  These surfaces will play a major role in 
this paper, they will be referred to as
holographic surfaces.

As there is no asymptotic region, the areas of these surfaces will
be generally finite\footnote{More specifically, the operators that
measure the areas of these surfaces will have finite (and generally
discrete) spectra.}. 
By the Bekenstein bound\cite{bek}, the state spaces
on which these boundary observables act must have finite dimension,
proportional to the exponential of  of the area, in Planck
units.  
In \cite{linking,hologr,superholo} it was shown that, at least
in the cases  of quantum general relativity and supergravity, with
a non-zero cosmological constant, this condition is met\footnote{For 
pure quantum general relativity, 
the constant of proportionality is not equal
exactly to $1/4$.  This apparently indicates that there is a finite
multiplicative renormalization of Newton's constant\cite{linking}.}.

In these theories the boundary state spaces are naturally
spaces of intertwiners (or conformal blocks) 
for $G_{q}$ on punctured surfaces, with the
punctures representing points where excitations of the quantum
geometry in the interior may end.  
It is these punctures which are identified with $D0$ branes.
To explain why, it is necessary to understood how dynamics is
formulated in this theory.

The main idea is that the dynamics is given in terms of a causal
histories framework, first proposed for quantum gravity in 
\cite{FM2}.  In this kind of theory, a history, $\cal M$,  is constructed
from a given initial state by a series of local evolution moves.  These
resemble the bubble evolution in general relativity and
as a result, these histories have a natural analogue of the causal
structure of general relativity. This makes it possible to identify
the analogues of light cones, horizons, spacelike surfaces and 
many-fingered time precisely at the quantum level, prior to any
continuum limit being taken\footnote{Indeed, these structures provide
the framework for formulating the existence of one or more classical
limits as a dynamical problem, analogous in some respects to 
non-equilibrium critical 
phenomena\cite{fmls1,stulee,janrenata,roumensameer}.}.

The dynamics is specified by the choice of local moves and the
amplitudes that are assigned to them.  Thus, in this kind of
theory a new kind of fusion between quantum theory and spacetime
is achieved in which states are identified with quantum geometries
that represent spacelike surfaces, and histories are both
sequences of states in a Hilbert space and discrete analogues
of the causal structures of classical spacetimes.  

The connection with string theory arises because 
small perturbations of states are 
parameterized
by closed loops drawn on the two surfaces ${\cal S}_{g}$ \cite{stringsfrom}.  
The small
perturbations of a history $\cal M$ are then in one to one correspondence
with the embeddings of certain time like combinatorial two dimensional surfaces
in $\cal M$.  As argued in \cite{stringsfrom}, if the theory has a
classical limit then the amplitudes for these perturbations must reproduce the
actions for excitations of string worldsheets.

When such an excitation takes place within a surface associated with
a holographic observer, it can end on one of that surface's punctures.
Thus, the punctures are structures on which strings end, which is to
say they are $D0$ branes.  The question is then to discover the observables
that describe the configurations of the $D0$ branes, and to deduce 
their effective dynamics.

The holographic principle, in the background independent form
proposed in \cite{tubes,screens}, plays a central role in achieving this.
There is a natural correspondence between bulk and boundary
states, which is suggested by topological quantum field 
theory\cite{louis-holo},
and confirmed in the case of quantum general 
relativity\cite{linking,hologr} and supergravity\cite{superholo}, with
a cosmological constant.  This is that the hamiltonian constraint,
which codes the spacetime diffeomorphism invariance, is satisfied
by the application of recoupling identities of quantum groups on
the quantum states of the theory. In a histories, rather than
a canonical theory, this must be imposed in a way which is
consistent with causality, so that information is not propagated
from the bulk to the boundary faster than the causal structure
of the quantum spacetimes allowed. Below, in section 3, we will see how
to do this\footnote{As discussed in \cite{hic}, this may resolve
several puzzles concerning the holographic hypothesis.}.

Once the map between bulk and boundary states is understood, the next
step is identifying the holographic observables that describe
the boundary states, and giving them a physical interpretation.
A clue for how to do this comes from Penrose's original work
on what he called the spin-geometry theorem\cite{roger-sn}.  This theorem
associates to generic, large,  spin networks with $N$ free ends 
an assignment of $N$ points on an $S^{2}$.  We show here that a
similar result is true in any dimension, and that this implies that
the observables which describe the punctures/$D0$ branes in the
present theory are closely related to those of a matrix model.
In the case that $G_{q}$ is $Spin(D)$ or an appropriate supersymmetric
extension of it we find that 
the observables associated with a surface with $N$ punctures then
corresponds to a matrix model in $D$ dimensions.  

This paper is divided into $11$ sections.  The next is devoted to an 
exposition of background independent membrane field 
theory\cite{tubes,pqtubes} in the causal histories
formulation of Markopoulou\cite{FM2}. Enough detail is given, and the
basic concepts are stressed, so that no prior knowledge of
loop quantum gravity need be assumed. 
 
Section 3 is then devoted to certain conceptual and technical
points regarding the treatment of gauge and diffeomorphism
invariance in this class of theories.  This is necessary
preparation for the main work of the paper, which is the
uncovering of the relationship to string theory. We begin this in 
section 4 where  we review the results of \cite{stringsfrom}
that show that small perturbations of the states are described
by loops embedded in the quantum geometry so that small perturbations
of causal histories are associated with embeddings of
combinatorial time like two surfaces in the histories. 
This leads us in section 5 
to the identification of punctures on the  
holographic surfaces
where  strings may end as $D0$ branes.  

In section 6 we identify
observables that code the dynamics of these punctures and show that 
they are closely related to those of the
matrix model.  We then  show that in $D=3$ there is a surprising
relationship between the observables that describe the $D0$ branes and Penrose's
original spin network formalism \cite{roger-sn}.  This suggests
a certain simplification of the operators that represent the
$D0$ brane dynamics.  

Finally, in section 7-10  we show that the bulk dynamics may be 
chosen so that the dynamics of the punctures which is induced by
the bulk-to-boundary map  
reduces, in the appropriate limits, that given by the matrix 
models. 
The bosonic and supersymmetric matrix models in $D=3$ are studied in
sections 7 and 8; in 9 and 10 we study their counterparts in
$D=9$.  
The main results of the paper are then summarized in the concluding
section. These are the basis of three conjectures we then state,
which concern the form of the background independent form of 
$\cal M$ theory.  
We also discuss there several implications of the
conjecture, which should be explored in future work.  The most
important of these is that there are consequences
for the $AdS/CFT$ conjecture\cite{juan-ads,witten-ads,other-ads}. 
If both that and the present
conjecture are true, than we can deduce the exact form of the
Hilbert spaces and operator algebras for the boundary conformal
field theories in $AdS$ spacetimes.  Similarly, there are
consequences for counting states of black holes.

\section{Background independent membrane field theory}

In this section we summarize the general structure
defined in \cite{tubes}, and the motivation coming from results in 
both string theory and 
in quantum general relativity and supergravity.
For more details the reader is referred to \cite{tubes,pqtubes,FM2}.

\subsection{The kinematics of background independent membranes}

We would like to describe a class of membrane theories in 
which the embedding space has no prior existence, but is instead
coded completely in the degrees of freedom that live on the two
surfaces.  As the background is not assumed to exist
a priori, there are no embedding coordinates.  Instead, 
the kinematical structure is given by the choice of a
quantum group or supergroup $G_{q}$ where $q$ will be assumed to be
taken at a root of unity.  $G_{q}$ has a finite list of
finite dimensional representations, which will be labeled
$j,k,l\ldots$.

Using standard constructions
from conformal field theory\cite{MS,witten-cs,louis2d3d,verlinde} 
or representation theory\cite{KL} there
is associated to each oriented closed two-surface
$\cal S$ of genus $g$ a finite dimensional space
${\cal V}^{g}$, which is called the space of conformal
blocks of WZW theory based on $G_{q}$.  The surface may
also be punctured, which means that there are 
a set of $N$ marked points, labeled by a set of representations,
$\{ j \} $ of $G_{q}$.
The same construction yields a finite dimensional vector
space ${\cal V}^{g,N}_{ \{ j \}}$.  When the manifold is
$S^{2}$, ${\cal V}_{\{ j \}}^{S^{2}}$ is called the space of
intertwiners.  

These spaces may be constructed in the natural
way described in \cite{MS,louis2d3d,verlinde} from 
${\cal V}^{S^{2}}_{jkl}$, which
is called the trinion.  An arbitrary surface is constructed by 
joining three punctured spheres at labeled circles, taking
the direct product of the ${\cal V}^{S^{2}}_{jkl}$ over all the
three-punctured spheres and summing over labels.
Every way of decomposing $\cal S$ in terms of trinions yields 
a basis of ${\cal V}^{\cal S}$, where the basis elements
are labeled by the circles and intertwiners. 
\begin{figure}
	\centerline{\mbox{\epsfig{file=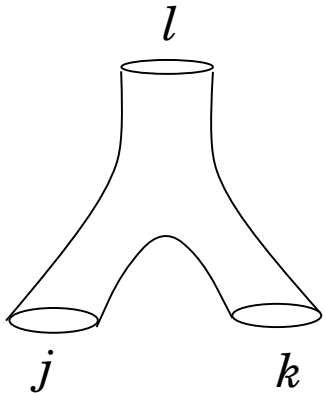,height=2in}}}
	\caption{A trinion which represents the intertwiner
	of three representations of $G_{q}$.}
	\label{trinionfig}
\end{figure}

The full state space of the theory is defined to be 
\f
{\cal H} = \bigoplus_{g} {\cal V}^{g}
\ff

As this is the fundamental assumption of this theory, let
us take a moment to stress it.  It means that the fundamental
things that the world is composed of are only 
systems of relations,  involving nothing but the
multiplication and decomposition of representations of some
fundamental algebra $G_{q}$.  At the fundamental level, spacetime
is nothing but a coarse grained description of processes in which
these systems of relations evolve. There is no background
geometry or topology, nor are there fields, particles, strings
or branes that move in them.  These must emerge
from the kinematics and dynamics of the states in $\cal H$.

\subsection{How quantum geometry is coded into conformal blocks}

Let us begin with geometry.  
It may seem that this arena is too poor to describe quantum
geometry.  However, this is not the case. There are in fact
bases of states in $\cal H$, whose elements
may be associated in an a natural way with quantum geometries of $d$
dimensions, for any $d$. 
The construction for $d=3$ is particularly simple\cite{tubes}, 
I discuss it in detail and then briefly mention describe the extension 
to larger 
dimension\footnote{This construction is relevant for the connection
to loop quantum gravity, but it is not used in the derivation of 
the matrix model and may be skipped by readers interested mainly in 
that result.}.

\subsection*{$d=3$ pseudomanifolds from conformal blocks}

\begin{figure}
	\centerline{\mbox{\epsfig{file=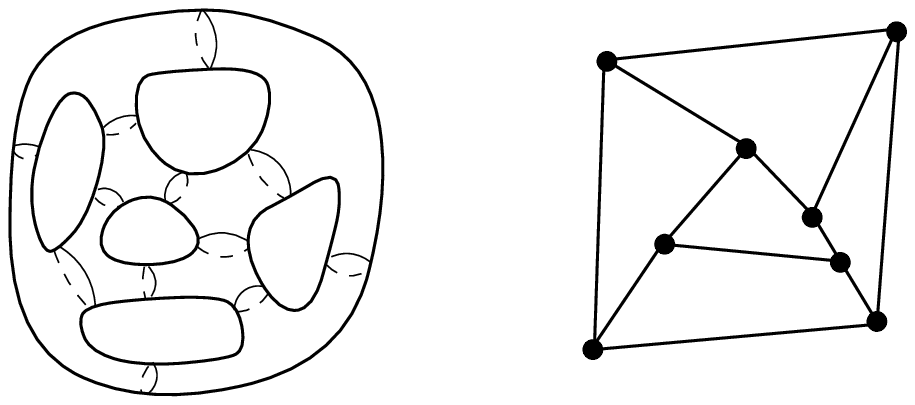,height=2in}}}
	\caption{A simple example of  two surface, and
	its dual graph $\Gamma$.}
	\label{tubenetfig}
\end{figure}

We consider, for  large
$g$, a decomposition of the genus $g$ surface, ${\cal S}^{g}$ 
into a number, $P$, of $4$-punctured
spheres.  We  label these $B_{d}^{i}$, with $i=1,\ldots,P$.  
The decomposition
is achieved by cutting ${\cal S}_{g}$ along a number, $N$, of circles
which we will label $c_\alpha$, with $\alpha =1,\ldots,N$.  We
can represent the decomposition by a $4$-valent framed graph $\Gamma$,
with $P$ nodes representing the $B_{d}^{i}$ and $N$ edges
representing the circles $c_\alpha$ (See Figure (\ref{tubenetfig}).).  
An associated basis
of states in ${\cal V}^{g}$ is given by assigning a representation
$j_a$ of $G_{q}$ to each circle $c_\alpha$ (or edge of
$\Gamma$) and assigning
a basis in the space of $4$ intertwiners $\mu_{i}$ to each 
$B_{g,d}^{i}$, (or node of $\Gamma$).  These basis elements
may then be labeled $|\Gamma, j_a, \mu_{i}>$.  Given
such a choice of basis for each $g$ yields a basis for
all of $\cal H$.

\begin{figure}
	\centerline{\mbox{\epsfig{file=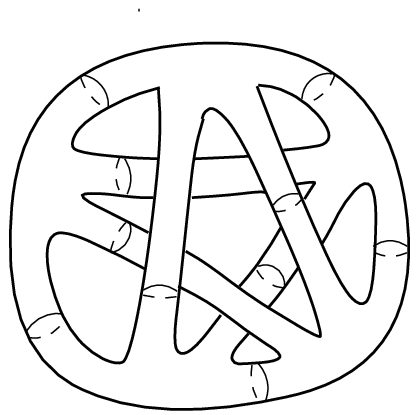,height=2in}}}
	\caption{A simple example of  two surface, divided into four
	punctured spheres.  The dual $3$-manifold is the boundary
	of the four-simplex.}
	\label{4simplexfig}
\end{figure}

Now to each element, $|\Gamma, j_a, \mu_{i}>$, 
of this basis we may associate a dual $3$
dimensional pseudomanifold ${\cal E}|\Gamma, j_a, \mu_{i}]$. 
Dual to each node of $\Gamma$ we construct a $3$ simplex,
with $2$ dimensional ``faces'' that correspond to the $4$
edges incident on it. Tying the faces together, following
the framing that follows from the orientation on the surfaces,
one constructs
a simplicial pseudomanifold of $3$ dimensions, whose faces are
labeled by the representations $j_a$ and whose 
simplices are labeled by intertwiners. (For a simple
example, see (\ref{4simplexfig}).)

The pseudomanifold ${\cal E}|\Gamma, j_a, \mu_{i}]$
inherits the labeling from the state $|\Gamma, j_a, \mu_{i}>$.
The faces are labeled by the representations $j_{\alpha}$
and the simplices by the intertwiners $\mu_{i}$.
We may give these a geometrical interpretation, which is
motivated by the results of quantum general relativity in
$3+1$ dimensions\cite{spain,sn1,sn2,vol2}. 
To each representation $j_{\alpha}$ we
may assign a quanta of $d-1$ dimensional area, given by
\f
A[j_{\alpha}] = l_{Planck}^{d-1}\sqrt{C[j_{\alpha}]}
\ff
where $C[j_{\alpha}]$ is the quadratic Casimir of
$G_{q}$.  This matches the result of the computation of
the area operator in quantum general relativity\cite{sn1,sn2} and
supergravity\cite{yilee1} in $3+1$ dimensions.  In those cases the operator
for the volume of a simplex was also computed and found to
be given by a finite, positive hermitian operator $\hat{Vol}$
on ${\cal V}_{j_1,j_{2},j_{3},j_{4}}^{S^{2}}$. 

For a general $G_{q}$ we must make choices of area and volume
operators, the first of which are functions of the representations, $j$
the second of which are positive operators on the 
${\cal V}_{j_1,j_{2},j_{3},j_{4}}^{S^{2}}$. While these may be
suggested by the results from canonical quantum gravity, there is no
need they be chosen so, and they may be chosen for a general $G_{q}$.
Once choices of the area and volume operators are made, 
the labeled pseudomanifold ${\cal E}|\Gamma, j_a, \mu_{i}]$ has
a geometrical interpretation in terms of volumes and areas.

The statement that the states in $\cal H$ are naturally associated
with pseudomanifolds means that the simplicial manifold
dual to a state $|\Gamma, j_a, \mu_{i}>$
may or may not be consistent with the application of the manifold
conditions on the simplices of dimension $0$ and $1$.  This means
that a general state $|\Gamma, j_a, \mu_{i}>$ can be
associated with a $3$ dimensional geometry, but one with 
defects of dimensions$0$ and $1$. In some cases these defects resemble
orientifolds, in that they may arise from a combinatorial manifold
by identification of lower dimensional simplices.  
When considered in isolation, these identified surfaces will 
behave as defects, inheriting dynamics from the fundamental
dynamics on $\cal H$.  These must be considered to be an inherent
part of the theory; I will argue in the conclusion
that at least in some cases the higher dimensional analogues of 
these may give rise to the $D$-branes.

We will also need to be able to talk about quantum geometries
with boundaries.  These are described by states that live in the
spaces of conformal blocks associated with 
punctured surfaces, which we denoted ${\cal V}^{g,m}_{j_a}$.   We may
also consider the space of all states with the same boundary,
given by\footnote{Sometimes the superscript
$m$ will be dropped when it is clear from the context.}
\f
{\cal H}^{m}_{j_a} = \bigoplus_{g} {\cal V}^{g,m}_{j_a} .
\ff

\subsection*{The $d=4$ case}

The key point in the above construction was to associate a $3$-simplex
as a region $\cal R$ of a two-surface $\Sigma$, with its connections to its $4$ 
neighbors across its $4$ triangles associated to a set of labeled punctures,
$c_{i}$ which, when cut, separate $\cal R$ from $\Sigma$.  For the 
case $d=3$ it is natural to take $\cal R$ to have genus $0$. The 
reason for this choice is that each puncture can be associated to a 
region of $\cal R$, each of which bounds all the others.  The meaning 
of this is that the dual skeleton of the surface of the three simplex 
can be drawn on $\cal R$ in such a way that its nodes are at the 
punctures and the edges that connect each pair of nodes are 
represented as lines that join the punctures.  We may note that these
lines may be themselves labeled with representations to give an 
intertwiner or conformal block on the four-punctured sphere.

When we go up to $d=4$ we want to represent $4-$simplices by regions
$\cal R$ of a two surface, with $5$ punctures, each labeling a 
$3$-simplex that is shared between this four simplex and its neighbors.
$\cal R$ should have thus have the topology of a $5$ punctured surface,
with the property that each puncture can be surrounded by a region of
$\cal R$, such that each of the five regions shares a border with each of the other 
four.  The minimal genus for $\cal R$  is then $1$ because the torus 
is the lowest genus on which five colors can be required for coloring 
a map.  

We may now repeat the above construction.  We need operators
to represent $3$-area and $4$-volume.  The former must be
a positive function of the representation labels, $j$, while the
latter must be a positive operator in 
${\cal V}^{g=1}_{j_{1},\ldots.,j_{5}}$ symmetric in all labels,
for each set $j_{1},\ldots,j_{5}$. 
Given $V$ we then label a basis of states
in each ${\cal V}^{g=1}_{j_{1},\ldots.,j_{5}}$ by a set of
eigenvalues $v^{i}$. (If these are degenerate
they can be supplemented by other geometric operators.). 

For $G_{q}$ related to $SO(4)$
or a supersymmetric extension of it, these operators can be derived from a 
canonical quantization of $4+1$ dimensional general relativity or
supergravity.  This has not yet been done, but there is no reason it
should not exist.  In any case, all that is required 
for the present construction is a choice of of area and volume 
operators for the group $g_{q}$

We now proceed to associate labeled $4$-pseudomanifolds to conformal
blocks of high genus surfaces by following the construction for $d=3$.  
Given a very large genus
surface,  ${\cal S}^{g}$  with $g$ chosen so it can be decomposed 
into a number  $P$, of $5$-punctured
torus's, which  are called $T^{i}$, with $i=1,\ldots,P$.  We fix
such a decomposition, which is achieved by cutting ${\cal S}_{g}$ along a number, 
$N$, of circles which we will label $c_{\alpha}$, with $\alpha =1,\ldots,N$. 
The sets of
eigenvalues $(v^{i},a^{\alpha})$,  then label a basis of states 
$|(v^{i},a^{\alpha}) >$, in
${\cal V}^{g}$.  To  each such state  we associate a labeled 
pseudomanifold by the following construction. We construct it from
$P$ $4$-simplices, $S^{i}$, each corresponding to a $T^{i}$.  Each is
attached to $5$ others along the circles
$c_{\alpha}$.  This gives us a $4$ dimensional pseudomanifold which is
associated to the decomposition (and hence the choice of basis in
${\cal V}^{g}$).  To each $4$-simplex we associate the four volume
given by the eigenvalue $v^{i}$ and to each $3$-simplex the three
area given by the eigenvalue $a_{\alpha}$. The result is a $4$ 
dimensional pseudomanifold, each of whose $3$ and $4$ dimensional
simplices has been assigned a volume.

The construction works in reverse as well, given a $4$ dimensional
pseudomanifold, whose $N$ $3$-simplices and $P$ $4$-simplices are 
labeled by eigenvalues
in the spectra of the $3$-area and $4$-volume operators, we can 
construct a two surface ${\cal S}^{g}$ for some $g$, 
by gluing together $P$ $5$-punctured tori. The correspondence between
labels on the pseudomanifolds and states in the spaces of intertwiners
on the $5$-punctured tori yields a state in ${\cal V}^{g}$.

\subsection{The general case}

To generalize the construction to general $d$ we must choose the
genus of the regions $\cal R$ such that it can be divided into $d+1$ 
regions, each of which shares a boundary with each of the others.
These will have at least genus $d-3$. Each of the  $d+1$ regions is
then punctured and the punctures on neighboring regions are joined.
To complete the construction one needs to find the appropriate volume
and area operators, presumably from a canonical quantization of 
supergravity in the appropriate dimension.

Finally, we may note that different bases are associated with different
geometrical decompositions of $\cal H$, of different dimension $d$.
This means that the dimension as well as the topological and metric
properties of the associated quantum geometry are dynamical quantities,
they are determined completely by the states. Which dimension,
topologies and geometries emerge in the classical limit is a dynamical
problem.  
We may also note that superpositions
of states that give rise to an interpretation in terms of a
$d$ dimensional quantum geometry will in certain cases be
interpretable as states associated with quantum geometries of
dimensions $d' \neq d$.  

These are all features that we would expect of a truly background
independent quantum theory of gravity.

\subsection{Reduction to quantum general relativity and supergravity}

Quantum general relativity and supergravity, with a cosmological 
constant, $\Lambda$, provide prototypes for the class of
theories just described.  In the Euclidean case, $G_{q} = SU(2)_{q}$
for general relativity and $G_{q} = SU(2|{\cal N})_{q}$ for supergravity,
for $\cal N$ at least up to $2$. The quantum deformation is given
by the cosmological constant, with $q=e^{2\pi \imath / k+2}$, 
with the level $k$ given by (\ref{cc}).   
The main results here are based on the existence of a 
closed form expression for the vacuum state, which is
given by\cite{kodama}, 
\f
\Psi_{CS}(A) = e^{{3 \over G^{2} \Lambda} \int Y(A)_{CS}}
\label{csstate}
\ff
where $Y(A)_{CS}$ is the Chern-Simons invariant
of the Sen-Ashtekar connection\cite{sen,abhay}.  A lot is
known about this sector of the 
theory\cite{BGP,jorgerodolfo}, including the fact
that the semiclassical behavior is that of small fluctuations
around deSitter or anti-deSitter spacetime\cite{kodama,chopinlee}.
The supersymmetric extensions are also known\cite{superstuff}.
An important fact is that the
observables algebra
of this theory requires that the spin networks be quantum
deformed\cite{qdeform}, as described in \cite{KL}. 
 
In the case of a spacetime with boundary, a class of boundary
conditions has been studied in which Chern-Simons theory is
induced on the boundary\cite{linking}.   With $\Lambda \neq 0$
there is a class of exact solutions to the quantum constraints,
labeled by the conformal blocks for all the possible choices
of punctures on the spatial boundary; the number of these
solutions is consistent with the Bekenstein bound\cite{linking}.

Similar results have been found also in the Lorentzian case\cite{hologr},
in which case $G^{q}= SU(2)_{q}\otimes SU(2)_{q}$, where, however
the representations and intertwiners are restricted by the
balanced condition $j_{left}=j_{right}$, first proposed 
in \cite{barrettcrane1}.

Finally, we note that when we take  $\Lambda \rightarrow 0$
this implies, by (\ref{cc}, that $q \rightarrow 1$
so that the states reduce to ordinary spin networks\cite{KL,qdeform}.  
Quantum general relativity in this limit has been studied in great 
detail\footnote{For a recent review of what is by now a large
literature, see \cite{carlo-review}.}
\cite{lp1,lp2,sn1,sn2,spain,carlo-review,BGP,jorgerodolfo,vol2},
and the results found have been verified by rigorous 
theorems\cite{rayner,chrisabhay,gangof5,thomas}. 

\subsection{Holographic observables}

There are several lines of argument that point to the conclusion that
in a quantum theory of gravity, all observables should be associated
with boundaries, of dimension, in the continuum limit,  
$1$ less than the spatial dimension.  These include the arguments
for the holographic hypothesis of `t Hooft\cite{thooft-holo}, 
whose relevance for
string theory has been argued by Susskind\cite{lenny-holo}.  We know from 
several examples that the holographic principle may be satisfied
in a quantum theory of gravity, although its exact formulation,
for cosmological theories, remains 
controversial\cite{lennywilly,garydon,hic,raphael}.

There are a parallel set of arguments in the literature on quantum
gravity and cosmology, put forward originally by Crane\cite{louis-holo},
and continued in \cite{pluralism,carlo-rel,lotc,FM3,FM4,cqc,screens}, to the 
effect that observables in a 
cosmological theory must be associated with a splitting of the universe
into two pieces.  According to this point of view, such a splitting may
correspond to a situation in which a subsystem of the universe,
delineated by the boundary, is studied by observers who are able
to make measurements only on the boundary.  One may then set up what
is called a relational\cite{carlo-rel} or 
pluralistic\cite{pluralism,lotc} quantum cosmology in which 
there is a space of states and an observable algebra associated to every
possible such spatial boundary.   These different state spaces are
tied together by linear maps whose structure is determined by
a functor from the category of cobordisms of $d-1$ dimensional
surfaces\cite{louis-holo,pluralism}.  This is done in such a way that 
the Hilbert
space associated to the whole universe is always one dimensional,
making it impossible to construct any non-trivial observables
associated to the universe as a whole. Given that no observers
have access to the entire universe, this is argued 
in \cite{louis-holo,carlo-rel,pluralism,lotc,FM3,FM4,cqc,screens}
to be a necessary property of a realistic quantum theory of 
cosmology\footnote{A general formulation of a cosmological holographic principle,
applicable to background independent theories, was given in
\cite{screens}.  As we showed there, that principle can be satisfied
in the class of theories defined in \cite{tubes} when the dynamics
satisfies certain restrictions.}.

In the present setting we can realize this approach to holographic
observables by associating algebras of observables with different ways
of splitting the two-surfaces into two parts, each representing different
halves of a quantum geometry, split along a boundary.
This is done
in the following way.  Given a quantum state, $|\Gamma, j_a, \mu_{i}>$ 
cut the surface ${\cal S}_{g[\Gamma]}$ along any subset of $N$ of its circles
$c_a$ which results in the splitting of ${\cal S}_{g}$ into
two pieces, which we will call ${\cal S}_{g}^{\pm}$.  The splitting
introduces an abstract surface, which may be taken to be a
punctured $S^{2}$, with $N$ punctures. These punctures inherit
the labels $j_a$ on the $N$ circles.  
\begin{figure}
	\centerline{\mbox{\epsfig{file=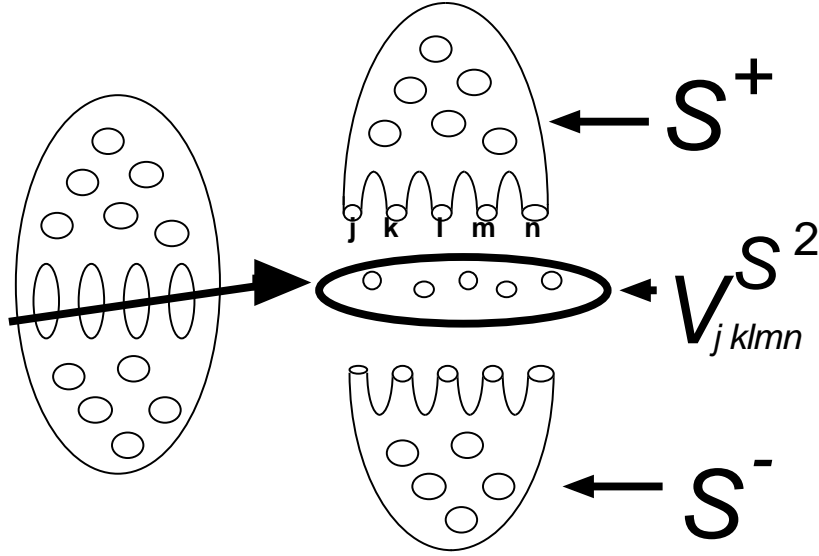,height=3in}}}
	\caption{A quantum geometry $|\Gamma, j_a, \mu_{i}>$ is cut along
	$5$ circles, yielding two bounded quantum geometries, ${\cal 
	S}_{g}^{\pm}$,
	which meet on a boundary which is represented by a $5$-punctured 
	$S^{2}$.}
	\label{cutfig}
\end{figure}

${\cal S}_{g}^{\pm}$ are each a quantum geometry with boundary constructed
from the graph dual to ${\cal S}_{g}^{\pm}$. It is interesting
to ask how the boundary is described in the picture
described above, in which there are bases of states
which have a dual description in terms of a labeled  $d$ dimensional
combinatorial pseudomanifold. In this picture, the
pseudomanifold must have a $d-1$ dimensional boundary,
$\partial {\cal S}_{g}^{\pm}$ which is 
composed of $N$ $d-1$ dimensional simplices, labeled
by the $j_a$.  In a continuum limit, this surface will 
correspond to a $d-1$
dimensional surface which divides space into two parts.

According to the holographic principle, the physical information that
observers living on the boundary 
$\partial {\cal S}_{g}^{\pm}$ may gain about ${\cal S}_{g}^{+}$
must then be represented in terms of a field theory on the boundary.
There is a state space naturally associated with that boundary,
which is ${\cal V}_{j_a}^{S^{2}}$. The algebra of observables on
${\cal V}_{j_a}^{S^{2}}$  then contains all the information that
observers may learn about the interior ${\cal S}_{g}^{+ }$ by
measurements made on the boundary.

This picture is realized  in 
quantum general relativity \cite{linking,pluralism,hologr},  
for the case of non-vanishing cosmological constant.
 
We may note that for finite numbers of punctures, the dimension
of the boundary state space ${\cal V}_{j_a}^{S^{2}}$ is finite,
even as $q \rightarrow 1$.
In quantum general relativity, this is consistent with the
Bekenstein bound, because the area of the boundary is
given by $l^2_{Pl}\sum_{\alpha}\sqrt{C(j_{\alpha})}$ \cite{linking}.   

Note that we have defined two different surfaces that
are associated with the boundary.  There is the $2$ dimensional
surface, which is the punctured $S^{2}$ on which the boundary
state space,  ${\cal V}_{j_a}^{S^{2}}$, lives.  Then there is
the $D-1$ dimensional pseudomanifold, $\partial {\cal S}_{g}^{\pm}$.
For $D=3$, these surfaces coincide, the latter gives a 
simplicial description of the punctured $S^{2}$, with a
puncture inside each triangular face.   However, for
$D>3$ they are different.  The construction of the
dual pseudomanifold will give us an embedding of the 
$S^{2}$ in the combinatorial pseudomanifold
$\partial {\cal S}_{g}^{\pm}$.  In the limit of a large
number of punctures, this must go over into the embedding
of a $2$-brane in the $D-1$ dimensional boundary.  This
$2$-brane is made of $0$-branes, which are the punctures
in the surface. This is the basis of the correspondence
to $\cal M$ theory we will develop below.  

To develop this picture we must have a clear understanding
of how the holographic principle is realized in this
background independent context.  
The main mystery of the holographic principle may be
stated in this context as follows.  The state space of a finite 
area boundary must
be finite dimensional, by the Bekenstein bound. However,
there seems to be an infinite dimensional space of possible
states for the fields in the interior.  The problem is how the
reduction from a potentially infinite number of bulk states
to a finite dimensional space of states, observable on the boundary,
is achieved.  
In the present case, the theory must provide a map $\Phi$
which reduces the infinite dimensional space of bulk states, 
$ {\cal H}^{N}_{j_a}$, to the finite dimensional space of
boundary states, ${\cal V}_{j_a}^{S^{2}}$.
There is a natural proposal for this map\cite{tubes,louis-holo} which
is motivated by the results of quantum general relativity
and supergravity with finite $\Lambda$ \cite{linking,hologr}.
This is that the map is given by Chern-Simons theory.  A state
$|\Gamma, j_a, \mu_{i}> \in  {\cal H}^{N}_{j_a}$
corresponds to a framed $G_{q}$ spin network with $N$
open ends, with labels $j_a$.  By standard 
constructions\cite{witten-cs,louis2d3d,KL}
it can be read as an $N$-intertwiner for $G^{q}$ and hence
as an element of ${\cal V}_{j_a}^{S^{2}}$.  We denote
the resulting linear map by
\f
\Phi :{\cal H}^{N}_{j_a} \rightarrow {\cal V}_{j_a}^{S^{2}}.
\label{holomap}
\ff
In these cases we will use the tilde to denote
\f
\Phi \circ |\Psi> \equiv  |\tilde{\Psi}> \in {\cal V}_{j_a}^{S^{2}}
\ff
This choice is motivated by results from quantum
general relativity and supergravity in $3+1$ dimensions.
In the canonical formalism of quantum gravity and
supergravity, the map $\Phi$ is is realized as the
loop transform \cite{lp2} of the Chern-Simons state (\ref{csstate})
in the presence of boundary states given by conformal blocks
of the punctured boundary.  The details of this are described
in \cite{linking}.  One way to say this is that there
is a sector of states in which the Hamiltonian constraint
of quantum general relativity (and supergravity)
is equivalent to the recoupling relations of framed
$SU(2)_q$ (or $SU(2|1)$) spin networks\cite{linking,qdeform,yilee1}.
What we are going to do is use the natural extension of this result
to {\it define} the relationship between the bulk and surface
states in $\cal M$ theory.

In a histories formalism of the
kind we will use here, the definition of the map will have to
be supplemented by giving time slicings 
of both the boundary and the interior.  We will be 
able to do this after we have
explained how histories are constructed in this theory.

\subsection{The form of the local evolution operator}

\begin{figure}
	\centerline{\mbox{\epsfig{file=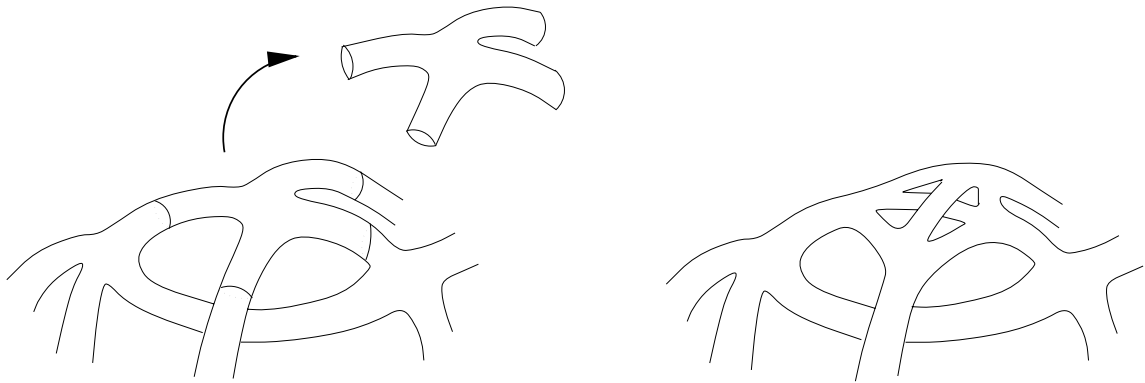,height=2in}}}
	\caption{An example of a substitution move.  Histories
	$\cal M$ are made up of successive applications of
	such moves.}
	\label{movefig}
\end{figure}

We now turn to the dynamics given to these background independent
membrane field theories.  For reasons given in \cite{tubes} we 
describe the dynamics in terms of a causal histories framework,
first proposed for quantum general relativity in \cite{FM2}.
In this approach the dynamics is given by a set of local
replacement moves. In each such move a specified  $m$- punctured surface
${\cal S}_{past}$ is removed from the surface $\cal S$.  This
results in $m$ free ends, with labels $j_{\alpha}$.  A new surface
with matching labels, 
${\cal S}_{future}$ is then inserted in its place.  
(See Figure (\ref{movefig}).)
This corresponds to the removal of a state $|past>$ in
${\cal H}^{m}_{j_a}$ and its replacement by a state
$|future>$.
Given a choice of operator $\hat{h}$ in each
${\cal V}_{j_a}^{S^{2}}$ and using $\Phi$ to associate
$|past>$ and  $|future>$ to states in ${\cal V}_{j_a}^{S^{2}}$, the amplitude
for each such move is given by 
\f
{\cal A}(move) = <\tilde{future}| \hat{h} |\tilde{past}> .
\ff

A causal history, $\cal M$ is specified by the choice of an initial
state, $|0> = |\Gamma, j_a, \mu_{i}> \in  {\cal H}$,
followed by a sequence $I=1,2,3,\ldots$ of substitution moves. 
The result is a
sequence of states, $|0>, |1>, \ldots,  |I>$, each
resulting from the previous one by a substitution move. 

The amplitude for a history is then given by the product
\f
{\cal A}[{\cal M}] = \prod_{I}{\cal A}(I)
\label{hamp}
\ff
The dynamics is then defined conventionally by summing
over all histories between given initial and final states.

As the substitution rules are local, this results in a structure
which is in many ways analogous to a continuous Lorentzian
spacetime\cite{FM2,tubes}.  Just as a classical spacetime 
may be constructed from a given
initial surface by a series of local, ``bubble evolution''
steps, in which the spatial hypersurface is evolved within
a compact region, a causal history $\cal M$ may be seen as
a spacetime structure generated by local evolution moves
on a state representing a quantum spacelike surface.

The precise way to describe this is to note that 
$\cal M$ has the structure of a causal 
set $\cal C$  \cite{FM2,tubes}\footnote{Recall that a causal 
set is a partially
ordered set with no closed causal curves 
that is locally finite. The application of causal sets to quantum
gravity was proposed first in \cite{rafael-poset,thooft-poset}.}.
To each of the regions removed or inserted by a substitution move,
one associates an
element $q \in {\cal C}$. A partial ordering may be defined
on $\cal C$ as follows: 
$p > q$ if there is a series of substitution moves that begins
with the removal of $q$ and ends with the insertion of $p$, such
that in each step the set removed overlaps the set inserted in
the previous move.  It is then easy to see that $\cal C$ has
the structure of a causal set.  We call it ${\cal C}[{\cal M}]$.

The causal set ${\cal C}[{\cal M}]$ gives to $\cal M$ discrete analogues of
many of the properties of Lorentzian spacetimes, such as 
many fingered time, light cones, and horizons\cite{FM2,tubes}. 
Various aspects of this are described in more detail 
in \cite{FM2,FM3,tubes,stulee}.  We may note that the association
of fundamental quantum histories with causal structures and
quantum states with quantum geometries makes this class of theories
generalizations of both  general relativity and quantum theory.
Each history $\cal M$, is {\it both} a sequence of states in a Hilbert
space, ${\cal H}$, defined without reference to any background
structure {\it and} a causal structure which shares many 
properties with a Lorentzian spacetime.

This ends the summary of the class of background independent membrane
theories proposed in \cite{tubes}. To summarize, a member of
this class of theories is given by the following choices:

\begin{itemize}
	\item{} 1)  The algebra or superalgebra, $G_{q}$.
	
	\item{} 2) The choice of substitution moves used in the evolution.
	
	\item{} 3)  The choice of the operators $h $
	that give the amplitudes for the causal histories $\cal M$.
\end{itemize}

\begin{figure}
	\centerline{\mbox{\epsfig{file=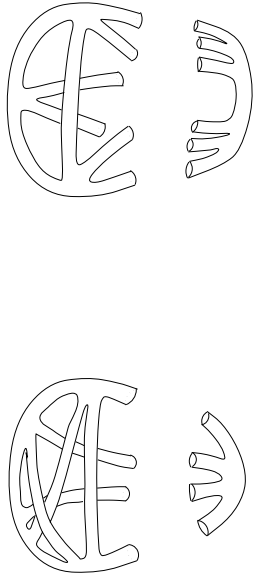,height=3in}}}
	\caption{Examples of past/future pairs that are used
	in the substitution moves in quantum general relativity.
	These come from cuts of the thickened four simplex.}
	\label{pastfuturefig}
\end{figure}

Particular  choices have been suggested for the theory corresponding
to quantum general 
relativity in the Euclidean \cite{mike,barrettcrane1} and
Lorentzian\cite{FM2} case.
(See Figure (\ref{pastfuturefig})).
The question that we may now address is whether there are any choices
that give continuum limits described by string theory.

\section{Consequences of a Planck scale causal structure}

There are several important issues we must
clarify, before we go on to study the possible relationship
with string theory.

\subsection{The continuum limit}

The most important physical question regarding the class
of theories we have just defined is whether 
there exists for any of them 
a continuum limit which reproduces classical general
relativity in some dimension, coupled to some set of quantum fields.
The main conjecture we are pursuing in this line of work is
that if any of these theories has a continuum limit, there must be
a consistent description of small perturbations around that limit,
and this in turn must
reproduce a string perturbation theory, because there is strong
evidence that all successful perturbative gravitational theories
are string theories.  The main question, to be discussed below,
is then how to construct observables of the fundamental theory
that will describe these string-like, perturbative degrees of
freedom, when the continuum limit exists.

We will return to the problem of the continuum limit
in the conclusions.  This is the main question that we
will {\it not} be able to solve in this paper.

\subsection{Spacetime diffeomorphisms, gauge fixing and discrete
causal structure}

To discuss the issue of gauge invariance we need some language
to describe a causal set.
Given the causal set
${\cal C}[{\cal M}]$ we can define an acausal set to be 
a set of elements $q_{i}\in {\cal C}[{\cal M}]$ such that
there are no causal relations amongst them. An antichain
$A$ is a maximal acausal set, which means that no 
$p  \in {\cal C}[{\cal M}]$ can be added to $A$ without violating
the condition that it be an acausal set\cite{FM3,FM4}.

A time slicing ${\cal T}=A(t)$ for $t=0,1,2,\ldots$ is a sequence of
antichains such that every $p \in {\cal C}[{\cal M}]$ is in some
$A(t)$.  This is just like a time slicing in general relativity,
except that the time parameter is discrete.

As in general relativity, there are many possible slicings
of a history.  These may be distinguished by gauge
conditions. Whenever we refer to a slicing we will have
to give conditions that pick it out.

This is especially important when we come to the description of
physical time evolution.
In general relativity the Hamiltonian on the physical, gauge
invariant, state
space is a function of the fields on the boundary.  To give
the correct interpretation of the present setting, which
insures that if there is a classical limit it will go over to
the correct description of evolution in the classical theory,
we must ask whether the local evolution moves are physical
or gauge. This is an issue because in general relativity
the Hamiltonian constraint generates both changes in the time
coordinate and local evolution moves.

In the kinds of theories defined in \cite{FM2,tubes} the local
evolution moves generate physical evolution because the
causal set ${\cal C}[{\cal M}]$ corresponds to the causal
structure of spacetime.  Thus the past set and future set
in a single local evolution move must correspond to physically
different sets of events, because they have different causal
relations with the other events, and this corresponds in the
classical theory to distinct physical spacetime events.  Another
way to say this is that the causal set 
associated to a history in general relativity is a spacetime
diffeomorphism invariant.  As such, the causal set
${\cal C}[{\cal M}]$ of a discrete quantum history is also
a gauge invariant description that, in the classical limit,
(if one exists) must correspond to the spacetime diffeomorphism
invariant causal set of a Lorentzian spacetime.

One consequence of this is that, unlike the case of state
sum models of topological quantum field theories, the sum over
histories is {\it NOT} a projection operator. This is because
there is no history where nothing happens to the states.
Every history is a sequence of real changes to the states
and these then correspond to a spacetime invariant description 
of real events.  

This does not mean that states must not satisfy additional
conditions, corresponding to the conditions on initial data
in the classical theory.  But these conditions are different
from the sum over histories, as each quantum history must
correspond, in the case that there is a classical limit, to
a spacetime diffeomorphism invariant description of a classical
history.

\subsection{Slicing conditions in the bulk and on the boundary}

There are special issues, in both the classical and quantum
theory, when a finite boundary has been introduced.
Note that to every antichain $A \in {\cal C}[{\cal M}]$ is
associated a state $|A> \in {\cal H}^{N}_{j_a}$.
Thus to a time slicing $A(t)$ is a sequence of states
$|A(t)> \in {\cal H}^{N}_{j_a}$. By 
(\ref{holomap}) this induces a sequence of states
$|\tilde{t}> \in {\cal V}_{j_a}^{S^{2}}$.

The question is then which slicing in the bulk is to correspond
to a choice of slicing on the boundary.  There is a unique
answer which is dictated by causality.  Given a history
$\cal M$ with boundary and 
a time slicing $\tilde{t}$ on the 
boundary, we define the {\it maximally past slicing}, 
$A(\tilde{t})$,
to be one that agrees with $\tilde{t}$ on the boundary
and is past inextendible in the interior.  This means that
for each slice $A(\tilde{t})$, which is an antichain in the
interior, there is no past replacement move, which would
remove a connected subset $a \subset A(\bar{t})$, disjoint from 
the boundary, and replace
it with elements $a^{\prime}$ such that,

1) $a^{\prime} < a$ in $\cal C$, and 

2) $( A(\bar{t})-a ) \cup a^{\prime}$ is an antichain.

Conceptually, such a slicing is one in which each spatial
slice is as close to the past lightcone of a given slice of 
the boundary as it can be and still be a complete
spacelike slice.  It is not difficult to argue that
for bounded histories $\cal M$ of the kind defined here,
given a slicing of the boundary, such a slicing of the
interior can always be chosen.

It is interesting to note that in the continuum limit the
slice approaches the past light cone of the cut of the boundary.
Thus, when $\Lambda <  0$ the limit that the boundary is
taken to infinity should reproduce the $AdS/CFT$ correspondence.
It is also intriguing to note that when $\Lambda = 0$ this
description must reproduce the heaven description of
Newman and collaborators\cite{tedn} in which solutions to
Einstein's equations are described completely in terms of
data on cuts of ${\cal I}^{+}$.

Given such a slicing we then have a sequence of states 
$|{t}> \in {\cal H}^{N}_{j_a}$.
By the holographic map,  eq. (\ref{holomap})
this will induce a sequence of states $|\tilde{t}>$ in the
boundary state space ${\cal V}_{j_a}^{S^{2}}$ given by
\f
|\tilde{t}> = \Phi \circ |t>
\ff

As a consequence, there must be a time evolution operator
$\hat{U}(t,s)$ on ${\cal V}_{j_a}^{S^{2}}$ such that
\f
|\tilde{t}> = \hat{U}(t,s) |\tilde{s}>
\ff

We will be concerned to identify this operator. To do so,
we need to identify a set of operators in 
${\cal V}_{j_a}^{S^{2}}$ which we may use to understand the dynamics
of the induced boundary states $| \tilde{t}>$.  
To do this it is helpful to have a physical interpretation for the
information in the boundary state space ${\cal V}_{j_a}^{S^{2}}$.
As we now show, a very interesting interpretation is suggested
by studying the properties of the small perturbations around
bounded histories.

\section{Small perturbations, and the identification of string
world sheets}

Now that we have cleared away some technical and conceptual
issues, we are ready to come to the heart of the matter, which is
the identification of small perturbations of a history $\cal M$
with string worldsheets embedded, in a suitable sense, in $\cal M$.

Let us begin by considering an arbitrary history and asking
what small perturbations look like.  A history is a
sequence of states $|I> = |\Gamma^{I}, j_a^{I}, \mu_{i}^{I}>$
each generated from the previous one by a local move.  Thus, a perturbed
history $|I'> = |\Gamma^{I'}, j_a^{I'}, \mu_{i}^{I'}>$
will involve a series of states  $|\Gamma^{I'}, j_a^{I'}, \mu_{i}^{I'}>$
each differing from the original by a small change.
We must then ask what are small changes of the states
$|\Gamma^{I}, j_a^{I}, \mu_{i}^{I}>$.  

This question was analyzed in \cite{stringsfrom}.  Here I summarize
the main ideas of that paper, which may be referred to for details.

Changes in the states are of two kinds, changes in the genus, $g$
and changes in the state in ${\cal V}_{g}$.  We will exclude the
first kind as the substitution moves generally change the genus;
we then look for changes that leave the genus fixed so that they
cannot be substitution moves.  These are then small changes in the
states
that cannot be confused with evolution moves.

To see what  small changes in  the states
are allowed that do not change the genus, let us pick
a basis constructed from a trinion decomposition of 
the surfaces ${\cal S}^{g}$.  The states $|I>$ are then
parameterized as $ |\Gamma^{I}, j_a^{I}, \mu_{i}^{I}>$,
where the circles $c_a$ arise from the trinion 
decomposition\footnote{Note that for $SU(2)$ the trivalent 
intertwiners are
unique, but this is not the case for larger groups.}.
Note that we are not allowed to change one of the $j_a$
arbitrarily, as the labels on the three circles of each
trinion must be such that there is at least one invariant
element of $j_{1}\otimes j_{2}\otimes j_{3}$.  In
the case of $SU(2)$ this condition yields the
triangle inequality, plus the condition that the sum
of the spins is an integer.  To satisfy these conditions one will 
in general have
to change two of the three labels of the circles bounding
each trinion, as well as the intertwiner.  
Thus we reach an important conclusion, which is that {\it the
small excitations of the states $ |\Gamma^{I}, j_a^{I}, \mu_{i}^{I}>$
are not local.}  Instead, the small excitations of
a state $ |\Gamma, j_a, \mu_{i}>$ are
constructed from closed loops $\gamma$ drawn on
the surface ${\cal S}^{g}[\Gamma ]$.  

To see this in more detail, note that given an elementary
representation $l$ and a closed loop $\gamma$ on ${\cal S}^{g}$
each state $ |\Psi> = |\Gamma, j_a, \mu_{i}>$ has a 
consistent perturbation given by 
\f
D_{\gamma}^{l }\circ |\Psi  > = |\Psi'> = 
|\Gamma, j_a', \mu_{i}'>
\label{consistent}
\ff
where $j_a' = j_{\alpha }\otimes l $ when
$\gamma$ intersects $c_a$ and is unchanged
otherwise, and an analogous condition holds for the intertwiners.

Given a fixed trinion decomposition, 
this defines the operator $D_{\gamma}^{l }$ on the space
${\cal V}^{g}$.

Now that we know what a consistent perturbation of a state is,
we can ask what a consistent perturbation of a history is. Taking
into account the discrete many fingered time of the causal histories,
we note that a consistent perturbation of a history must give 
a consistent perturbation of every state that may be obtained by
slicing the history to produce a maximal acausal set.  As shown in
\cite{FM2,tubes}, every such slicing is associated with some
state in $\cal H$.  A consistent perturbation will be one that
is a linear combination of perturbations of the form of 
(\ref{consistent}) for every possible slice.  This will be true if
it gives a loop $\gamma_{\Gamma}$ for every slice of
$\cal M$ that produces a state of the form of a linear combination
of the basis states
$ |\Psi> = |\Gamma, j_a, \mu_{i}>$ associated with some
trinion decomposition.

Furthermore, we require that the perturbation be causal, which means
that in any sequence of states, $|I>$  that defines a history,
$\cal M$, the changed labels in $|J>$ are in the future of the
changed labels in $|I>$ whenever $J>I$.  

As argued in \cite{stringsfrom} the result is that consistent
perturbations of a history $\cal M$ are parameterized by a choice of
a fundamental representation $l$ of $G^{q}$ and an embedding
of a timelike (with respect to the induced causal 
structure $\cal C$ of $\cal M$) surface $\Delta$ in $\cal M$.  
The change in amplitude of the perturbed history can then
be expressed as a spin system on this surface, whose couplings
are induced from the amplitudes of the fundamental histories
and depend both on the choice of evolution moves and their
amplitudes and the embedding.  Details are given in \cite{stringsfrom}.

However, even without giving any details we can make the following
argument.  Suppose that the history $\cal M$ is semiclassical
so that it is a critical point of the path integral and represents
in a suitable limit a classical manifold. Then the degrees of
freedom of the perturbed
histories must contain the massless modes associated with small
perturbations around the classical limit.  These must be
carried by the induced worldsheet theory, since that codes the
small perturbations around any history. This means that the
effective theory defined on the worldsheet must carry the
massless spin 2 degrees of freedom, as we know these 
must be present if the theory, as assumed, has a classical
limit dominated by $\cal M$.  But this means that in the continuum
limit that worldsheet theory must reproduce the action for a 
critical string theory. 

Hence, the original non-perturbative theory must be a background
independent formulation of string theory, in the sense that the
theory of small perturbations around the classical limit reproduces
perturbative string theory. In fact, one may also argue that the 
leading term in the change in
the amplitude of the perturbed history is proportional to the
induced area of the time like two surface $\Delta$, and so matches
the Nambu action\cite{stringsfrom}.
 
\section{The identification of punctures as $D0$ branes}

The string world sheets we have identified must be closed, 
unless they end on boundaries.  It is natural to identify
$D0$ branes as points on boundaries where strings 
end\footnote{It might be objected that these are not 
necessarily $BPS$ states,
and so do not share all the properties of $D$-branes in string theory.
It might be better to say that these are objects that will behave
as $D0$ branes in the appropriate dynamical setting.  To avoid
inventing a new terminology, we will simply call them $D0$ branes.}.  
A setting that
permits this is the following.  Let us fix an $N$-punctured surface,
${\cal S}_{j_a}$, which we may for simplicity assume
is an $S^{2}$.  Let us consider a piece of a history consisting
of a sequence of states $|I>=|0>,|1>,|2>,\ldots$, generated by
local moves from an initial state $|0>$, which are chosen to live in the space 
${\cal H}^{N}_{j_a}$ with fixed boundary, which matches
${\cal S}_{j_a}$.  We will restrict attention to histories
in which the local moves act only in the interior of the quantum
geometries of the states $|I>$, so that each state in the sequence
has a boundary in the same space ${\cal S}_{j_a}$.
These may be considered to represent a piece of a causal
history in which we have restricted the evolution on the boundary
so as to not change the number and labels of 
the punctures\footnote{Boundary conditions that 
realize this condition in general relativity
and supergravity are described in \cite{linking,hologr}.}.

Let us denote a particular such history by ${\cal M}^{+}$, where
the $+$ indicates that it is half of a compact history, and
so may be joined by another piece ${\cal M}^{-}$ such that
each of the states in ${\cal M}^{+}$ matches one in 
${\cal M}^{-}$ by being joined on ${\cal S}_{j_a}$.
Now small perturbations of ${\cal M}^{+}$ correspond to choices
of fundamental representations and  loops drawn on
each state, $|I>$.  However, we may note that it is no longer
necessary that the loops be closed, instead they may end on
the punctures of the ${\cal S}_{j_a}$.  This identifies
these punctures as $D0$ branes, i.e. as zero dimensional spatial
objects on which strings can end.

There are other similarities between the punctures and $D0$ branes.
If a punctures label $j_a$ is one of the fundamental representations,
than only one string may end on it.  However if $j_a$ is
a larger representation, than as many strings can end on it as
there can be products of the fundamental representations that
have $j_a$ in their decomposition.  As a result,
punctures can be combined according to the laws of multiplication
of representations of $G^{q}$.  

For example, in the case $G_{q}=SU(2)_{q}$ the smallest punctures
are associated with $j=1/2$, and only one string can end on
each one.  The symmetry group associated with a puncture is
consequently $SU(2)$. But $n$ strings can end on a puncture
of spin $n/2$.  The associated symmetry group is $SU(n+1)$.
Thus by combining punctures one raises the symmetry from
$SU(2)^{n}$ to $SU(n+1)$

To see what the effect of this should be in a semiclassical
description, we must look ``inside'' each representation,
at the behavior of the classical object represented.  For
example, a state in the vector representation must correspond
to a vector in $D$ dimensional space. Small perturbations
then correspond to small motions of the classical object.  
In the case of $SU(2)$, linearization around a vector
gives $SO(2)$, while linearization around a spinor
gives $U(1)$.  

In the linearized approximation, the representation 
labels associated with the $D0$ branes become
charges which generate the subgroup that preserves the
classical object.  In this approximation, the products
of representations becomes the addition of
abelian charges that characterize, in the semiclassical
theory, the superpositions of
$D0$ branes.
In our example, the combination of $n$ punctures then induces,
in the linearized approximation, a symmetry enhancement from
$U(1)^{n}$ to $U(n)$. This 
is in agreement with what happens when $n$ D-branes
are combined in the semiclassical picture.  

We then hypothesize that in the continuum limit the
punctures are associated with $D0$ branes.  The rest of this
paper is devoted to exploring the consequences of this identification.

\section{Punctures and matrix models}

The first problem is to find the operators that, acting
on the boundary state space ${\cal V}_{j_a}^{S^{2}}$,
describe the dynamics of the punctures.
We begin with the observation that 
changes in ${\cal V}_{j_a}^{S^{2}}$ 
correspond to change in an intertwiner, with fixed puncture
labels.  What are the most local observables which measure these
changes?  A complete, in fact overcomplete, set of observables
in ${\cal V}_{j_a}^{S^{2}}$ correspond to Wilson loops, taken
in the fundamental representation of $G$:
\f
T[\gamma ] = Tr\left [  e^{\int_{\gamma}A} \right ]
\ff
where $\gamma$ is a non-intersecting loop in the
punctured boundary ${\cal S}^{2} - \{ {\alpha} \}$ and $A$
is the flat connection on the punctured surface
that defines the phase space
whose quantization gives ${\cal V}_{j_a}^{S^{2}}$

The simplest of the loops in ${\cal V}_{j_a}^{S^{2}}$
are those that surround a single puncture. However, these
have fixed values given by $j_a$, so they contain 
information that does not vary in time.  In quantum general
relativity with appropriate boundary conditions, these correspond
to the areas of the regions of the boundary containing
the punctures, by fixing the boundary as we have we have
in essence fixed the areas of the regions of the boundary.

The next simplest observables correspond to loops
$\gamma_{ab}$ that surrounds two punctures, $a$
and $b$.  To define these we fix a base point 
$p \in {\cal S}^{2} - \{ {\alpha} \}$ and define loops
$\gamma_{a}$ based at $p$ that sorround a single puncture
$a$. We then define composite loops 
$\gamma_{ab}= \gamma_{b}\cdot \gamma_{a}$.
We then have a matrix of operators
\f
\hat{T}_{ab} \equiv \hat{T}_{\gamma_{ab}}
\ff
The expectation values of these operators will, in fact,
contain all the observable information concerning the 
states in the boundary Hilbert space, ${\cal V}_{j_a}^{S^{2}}$.
They encode all the information about the configurations of the
punctures. As there is no fixed background geometry, this information
must be coded, as it is here, in relational variables that
give information about relationships between pairs of punctures.
\begin{figure}
	\centerline{\mbox{\epsfig{file=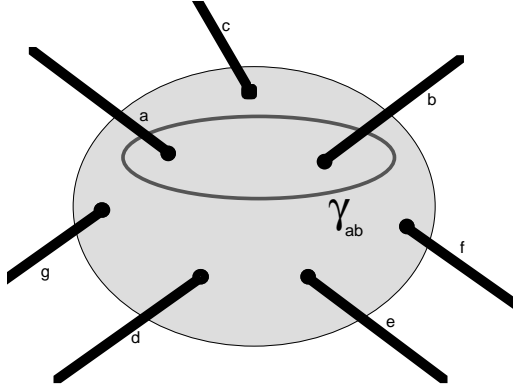,height=2in}}}
	\caption{ The $\hat{T}_{ab}$ operator. The shaded region is the
	punctured boundary  ${\cal S}^{2} - \{ {\alpha} \}$.}
	\label{Tabfig}
\end{figure}
But if, as we have argued, the punctures correspond to 
$D0$ branes, then, for an appropriate choice of
$G_{q}$, this matrix of operators should correspond
to the matrix description of the dynamics of
$D0$ branes, hypothesized by Banks, Fischler, Shenker and
Susskind.  In that description there is a matrix
of operators $\hat{X}^{i}_{ab}$ for every transverse
coordinate $i$ in light cone coordinates of flat
$11$ dimensional Minkowski spacetime.  

We will first establish a correspondence for any odd dimension $D$.
For a $D$ dimensional matrix model we have a matrix of
coordinates $\hat{X}_{ab}^{i}$, with $i=1,\ldots,D$.  
To establish a correspondence we have to first choose
the algebra $G_{q}$ appropriately and then find out where
this transverse coordinate $i$ lives.  
As there
is no background manifold there is no transverse space,
this has to emerge in the appropriate continuum limit.

In odd dimension $D$ we can represent the bosonic matrix model by
choosing  
$G_{q}=Spin(D)$. We will take 
traces of holonomies in a spinor representation of
$Spin(D)$ in which there are $D$ gamma matrices $\Gamma^{i}$.
We may then
define the classical phase space functions,
\f
T^{i}[\gamma ] = Tr \left [ \Gamma^{i} e^{\int_{\gamma}A} \right ]
\label{indexin}
\ff
From (\ref{indexin}) we have a matrix of quantities given by
\f
T^{i}_{ab}= Tr \left [ \Gamma^{i} e^{\int_{\gamma_{ab}}A} \right ]
\label{matrix17}
\ff
We make several comments on these variables.
First, note that  the diagonal elements use
$\gamma_{ab}=\gamma_{aa}$, the curve that wraps twice around the 
single puncture.  

Second, note that as in the matrix model, we can consider the
$N$ eigenvalues $r^{i}_{a}$ of the $D$, $N\times N$ matrixes
$T^{i}_{ab}$.  These give us $N$ points in $R^{D}$.  In the 
limit  $N\rightarrow \infty$ the usual arguments 
from the original matrix mode\cite{dWHN,BFSS} may be used to define
a mapping from an $S^{2}$ into
$R^{D}$.  Thus, in this limit we see emerge a description of
a membrane moving in $R^{D}$.  

However there are some differences with the
usual matrix model. First, there are constraints coming from the 
conditions that the trace of the holonomy around single punctures
are fixed.   The effect of these will depend on the dimension.
Then we must notice that the $T^{i}_{ab}$ are  
not a commuting set, instead $\{ T^{i}_{ab},T^{j}_{cd}\} \approx 1/k $
whenever $a$ or $b$ is equal to $c$ or $d$. Thus
the quantities (\ref{matrix17}) cannot correspond directly to the
coordinates of a matrix model. However, 
recall that in $2+1$ and $3+1$ dimensions the
cosmological contant is inversely proportional to a power of
$k$, so that $\Lambda \rightarrow 0$ when $k \rightarrow \infty$
(see (\ref{cc}).  
This suggests that we should seek to recover the standard matrix
model, which corresponds to supergravity in $11D$ with
vanishing cosmological constant, in the limit $k \rightarrow \infty$.
We will then need to rescale the fields so as
to recover the matrix model in the limit $k \rightarrow \infty$.

Finally, we must also note that the quantities (\ref{matrix17}) are 
not gauge invariant.
They are non-vanishing when evaluated on the classical
phase space, which is the flat connections on the
punctured surface.  If we turn them into operators in the 
space of intertwiners
they will vanish, because only gauge invariant quantities are
non-vanishing.  This is because we are constructing the theory
in terms of gauge invariant quantities; and any particular
coordinates on a background manifold which emerges from this
description will be gauge non-invariant.  
For this reason we will proceed to first  construct 
formal operators that represent pieces of gauge invariant
quantities, which include the  $T^{i}[\gamma ]$.   We then
combine these to form gauge invariant operators.

\begin{figure}
	\centerline{\mbox{\epsfig{file=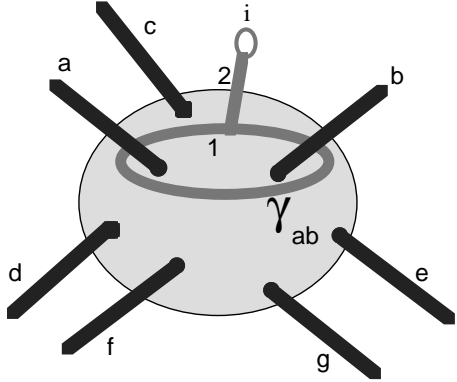,height=2in}}}
	\caption{ The $\hat{T}_{ab}^{i}$ operator. The 
	$i$ in the circle stands in the classical theory for
	the trace of the phase space observable against $\Gamma^{i}$.
	In the quantum theory it stands for a formal operator in which
	the $i$ is contracted against another open $2$ line that
	acts somewhere else
	in the state.}
	\label{Tabifig}
\end{figure}

One of the gauge invariant quanties we can construct is closely
related to the
hamiltonian of the $D$ dimensional matrix model. To do this first find
a set of momentum functionals $P_{i}^{ab}$ so that,
\f
\{ T^{i}_{ab}, P^{cd}_{j} \} = \delta^{i}_{j}\delta^{c}_{a}
\delta^{d}_{b}
\ff
These must exist, but we will not construct them explicitly, 
instead, the appropriate quantum
operators will be constructed below.  
We may then construct a classical hamiltonian that matches that of the bosonic
sector of the matrix model, in the limit $k \rightarrow \infty$ in
which the $T^{i}_{ab}$'s become commuting variables.
\f
H= { g^{2 } \over 2} Tr \left [ P_{i}P_{i} \right ]
+ {1 \over g^{2}} Tr \left ( [T^{i},T^{j}],[T_{i}, T_{j}]
\right )
\label{classH}
\ff
where $g$ is a dimensionless coupling constant. 

What this means is that a system which is closely related to the matrix
model may be coded in the classical phase space of the
boundary observables.  However, this is only a step,
for in the present theory the dynamics has already been
given by the substitution moves in the bulk.  The dynamics of the
punctures will then be induced by the bulk-to-boundary map, applied
as discussed in section 3.  What we then
want to know is whether the bulk evolution moves can be chosen so that 
the dynamics of the matrix model is {\it induced} as
an effective quantum dynamics of the punctures.
To do this we will proceed in 
two steps.  First, we construct a quantum hamiltonian, $\hat{H}$, 
on the boundary,
state space ${\cal V}_{j_a}^{S^{2}}$
which yields (\ref{classH}) to leading order.
Second, we will find a set of substitution
moves and amplitudes which reproduce the quantum
dynamics generated by $\hat{H}$.

We will carry this out  first for the bosonic 
matrix models in $D=3$, then we extend the correspondence, in turn to 
the supersymmetric matrix model in $D=3$ the bosonic matrix model
in $D=9$, and, finally, a supersymmetric matrix model in
$D=9$ that reduces in a certain limit to the dWHN-BFSS model.

\section{The $D=3$ matrix model}

We first study the bosonic matrix model in $D=3$ in which case
we take $G_{q}= SU_{q}(2)$.  
 
We begin by asking how the classical limit should emerge.
As the punctures must correspond, in the reduction to
quantum general relativity, to quanta of area, the continuum
limit must be the limit of large area, in Planck units.  In
the cases associated with general relativity or supergravity
in $3+1$ dimensions,in which $G= SU (2)$ or $Osp(1|2)$,it is known
that the area of a surface with $N$ punctures with labels
$j_{\alpha}$ is proportional to $\sum_{\alpha}\sqrt{C(j_{\alpha})}$
where $C(j_{\alpha})$ is the quadratic casimer operator.  Furthermore, 
as shown in \cite{linking} the most probable way of
reaching the limit of large area, in which the Bekenstein
bound is saturated, is the one in which each puncture
carries a minimum area, corresponding to the smallest representation,
$j_{1/2}$ of $G_{q}$. In this case the dimension of the boundary state
space saturates the Bekenstein bound as the area is taken to infinity.
This tells us that we should expect the continuum limit
to emerge in the case that all the punctures  have the
same value $j_{\alpha}=j_{1/2}$.  
For these, most probably states,  the limit of large area is 
$N \rightarrow \infty$, where $N$ is the number of punctures.

\subsection{Matrix valued operators on 
${\cal V}^{{\cal S}^{2}}_{j_{\alpha}}$.}

In the case of $D=3$, the  the diagonal elements
$T_{aa}^{i}\equiv v^{i}$ are fixed by the condition that the
Wilson loops around single punctures are fixed in terms of
$k$ and the representation label.  
Physically, this means that the  vectors $v^{i}_{a}$ live in
$S^{2}$. This is reminiscent of 
the spin geometry theorem, which was the motivation for the
original introduction of the spin network formalism by 
Penrose in the early 1960's\cite{roger-sn}.  There a map is found between
$SU(2)$ spin networks with $N$ ends, labeled by spins
$j_{\alpha}$ and $N$ points on a two sphere.  This is gotten
by constructing an operator $\hat{X}_{ab}$ which connects
the $a$'th to the $b$'th external edge with a spin $1$ line.  
If $|\Gamma >$ is an ordinary 
$SU(2)$ spin network, with $N$ ends, then the map $\Phi$ produces in this
case an $SU(2)$ intertwiner $|\tilde{\Gamma}>$\footnote{The application 
of $\Phi$ is equivalent to the operation Penrose
called the evaluation of a spin network.}.
Since
$\hat{X}_{ab}$ acts adjacent to the ends, its action on
intertwiners $|\tilde{\Gamma}>$ is well defined.   Penrose
then finds that the expectation values,
\f
Y_{ab}\equiv  { <\tilde{\Gamma} | \hat{X}_{ab}|\tilde{\Gamma} > \over
<\tilde{\Gamma}|\tilde{\Gamma} >} = \cos(\theta_{ab})
\ff
for $a \neq b$ define $N(N-1)/2$ angles $\theta_{ab}$ which,
in the case that they do not vanish,
may be interpreted consistently to give relative angles
between $N$ points in a two sphere\cite{roger-sn}.  This is called
the spin-geometry theorem.  

In fact, it is easy to show that for $G=SU(2)_{q}$,
\f
\hat{T}_{ab}= {16 \pi^2 j_{a} j_{b} \over k^2} \hat{X}_{ab} 
\ff

\begin{figure}
	\centerline{\mbox{\epsfig{file=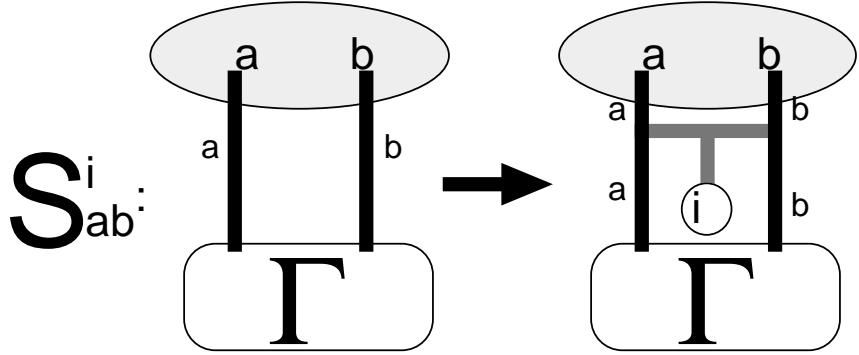,height=2in}}}
	\caption{The definition of the operator $\hat{S}^{i}_{ab}$
	in the $D=3$ cases.
	The added trinions are in grey, all the labels are in the 
	adjoint representation.  The remainder of the surface is
	indicated by $\Gamma$, other ends on the boundary are not
	shown.}
	\label{Sabifig}
\end{figure}

Thus, the $G=SU(2)$ case of the construction of the previous
section is closely related to Penrose's spin geometry theorem.
This correspondence suggests the following construction, which 
we will shortly extend to general $G_{q}$.
We introduce a  
set of formal operators $\hat{S}^{i}_{ab}$ that act  on the
bulk  state space, ${\cal H}^{N}_{j_a}$.
Given a state 
$ |\Gamma, j_a, \mu_{i}> \in {\cal H}^{N}_{j_a}$, 
these act as follows. (See Figure (\ref{Sabifig}).)

We break the tube going to the puncture $a$ along a circle $c_{a}$ with
label $j_{a}$. We then insert a new trinion as follows.  
Two of the three punctures
of the trinion join the two sides separated at $c_{a}$, with identical
labels $j_{a}$.  The third we label with the adjoint representation
$Adj$ of $G_{q}$ and leave open for the moment.  We then do the
same with the puncture $b$.  We then introduce a third trinion
one puncture of which is tied to the free puncture of the new
trinion inserted adjacent to $a$. A second puncture is joined to the
free puncture  adjacent to $b$.  Each of these have labels
$Adj$. At the third puncture, we introduce an operator that
corresponds to tracing over the gamma matrix $\Gamma^{i}$.  
This is a formal operation, it means that this operator stands
for a family of operators where the $i$ index is contracted
with another $i$ index, with a tube labeled by $Adj$.  
We call the resulting formal operator $\hat{S}_{ab}^{i}$.  
 
This then induces an operator in ${\cal V}_{j_a}^{S^{2}}$,
which we call $\tilde{S}_{ab}^{i}$, defined with the help 
of the holographic map (\ref{holomap}). Given
$|\Psi > \in {\cal V}_{j_a}^{S^{2}}$ let $|\Psi^{\&}>$ be
any state in ${\cal H}^{N}_{j_a}$ such that
\f
\Phi \circ |\Psi^{\&}> = |\Psi>
\ff
This is non-unique but it doesn't matter,  because we are about
to act only on the edges that are adjacent to the punctures,
and these are fixed.  Then we define the action of $\tilde{S}_{ab}^{i}$
on $|\Psi >$ by 
\f
\tilde{S}_{ab}^{i} \circ |\Psi > \equiv \Phi \circ {S}_{ab}^{i} \circ 
|\Psi^{\&}>.  
\label{Sdef}
\ff
 
It is also straightforward to show that $\tilde{S}_{ab}^{i}$
is related to $T^{i}_{ab}$ in the $SU(2)_{1}$ case by 
\f
T^{i}_{ab}= {16 \pi^2 j_{a} j_{b}\over k^2}  \tilde{S}^{i}_{ab} 
\label{scaling}
\ff

The spin geometry theorem then gives us some insight into the meaning of 
the 
off diagonal elements of the $T_{ab}$ and $T_{ab}^{i}$ at
least for $D=3$. When treated as classical phase space
variables, the combination 
\f
Y_{ab}^{AB}=\epsilon^{AB} 
T_{ab}+ \imath \tau^{AB}_{i}T^{i}_{ab} 
\label{notscaling}
\ff
may be interpreted as an $SU(2)$ element that rotates the point
$v^{i}_{a}$ into the point $v^{i}_{b}$ on the $S^{2}$.  One can
see from this that
the off diagonal elements of $T^{i}_{ab}$ are proportional to
the $\sin(\theta_{ab})$, thus they measure the distance on the
$S^{2}$ between the 
the points $v^{i}_{a}$ and $v^{i}_{b}$.

However, we may also note from the definition of the operators
(\ref{indexin}) and (\ref{Sdef}) that the $ab$ elements
of $<\Gamma | T^{i}_{ab}|\Gamma >$
will vanish when the edges $a$ and $b$ come from disconnected
pieces of the spin network $|\Gamma >$.  (As this is a formal
operator it means it vanishes when the index $i$ is contracted
against any operator.)  This means that the off diagonal elements
vanish when the ends corresponding to the points fall into
disconnected clusters, corresponding to $\Gamma$ having
disconnected pieces.  

Thus, a potential of the form of that in (\ref{classH}) is
minimized either by the points lying on top of each other,
or falling into disconnected clusters.  

\subsection{Construction of the quantum matrix model Hamiltonian on
the space of intertwiners}

We now are ready to express the matrix model dynamics, given
classically by (\ref{classH}),  in terms of
operators on ${\cal V}_{j_a}^{S^{2}}$.  Care must be taken
as $\hat{S}^{i}_{ab}$ differs from $T^{i}_{ab}$ by powers
of $k$ and $k$ is taken to infinity in the limit that defines
the vanishing of the cosmological constant and hence the
limit in which the matrix model is recovered.
We know that it is the 
$T^{i}_{ab}$'s that must be the basic operators that define
the matrix model, as it is these that go
into the commuting coordinates of the matrix theory in the limit
that   $k \rightarrow \infty$.  But
when computing it is easier to work with the  $\tilde{S}_{ab}^{i}$ 
as they have a simple action on states.    

We may use (\ref{scaling}) to write a quantization of the potential
energy term of
classical matrix model hamiltonian, (\ref{classH}), as an
operator on  ${\cal V}_{j_a}^{S^{2}}$ 
\f
\hat{V}=  {1 \over g^{2}} 
\left ( { 2 \pi \over k} \right )^{8}
Tr \left ( [\tilde{S}^{i},\tilde{S}^{j}]
[\tilde{S}_{i}, \tilde{S}_{j}]
\right ).
\ff
To define the kinetic energy term we need to define the
conjugate momentum operator,
  $\tilde{\pi}^{ab}_{i}$   such that
\f
\{ \tilde{S}_{ab}^{i}, \tilde{\pi}^{cd}_{j} \} = \delta^{i}_{j}\delta^{c}_{a}
\delta^{d}_{b}
\ff
This is defined diagramatically below in Figure (\ref{piabifig}).
We may then write the hamiltonian corresponding to the matrix model.  
\f
\tilde{H}= { g^{2 } k^{4} \over 32 \pi^{2}} 
Tr \left [ \tilde{\pi}^{i}\tilde{\pi}^{i} \right ]
+ {1 \over g^{2}} 
\left ( { 2 \pi \over k} \right )^{8}
Tr \left ( [\tilde{S}^{i},\tilde{S}^{j}]
[\tilde{S}_{i}, \tilde{S}_{j}]
\right ).
\label{qH}
\ff
Here we have used the fact that all the punctures are
taken in the fundamental representation $j=1/2$, which is
necessary to have the
largest entropy per area of the holographic surface.

\subsection{Realizing the $D=3$ Matrix theory hamiltonian in terms of 
local evolution moves}

We now proceed to show that there  is a choice of the 
fundamental dynamics that reproduce the
Hamiltonian of the $D=3$ matrix model acting on the boundary
state space ${\cal V}_{j_a}^{S^{2}}$.
The potential energy term in  
(\ref{qH}), is formally given by,
\f
\hat{V}=  Tr\left ( [\tilde{S}^{i},\tilde{S}^{j}]
[\tilde{S}_{i},\tilde{S}_{j}] \right )
\label{hatU}
\ff
We may note that the $\tilde{S}_{ab}^{i}$ operators, like
the $\hat{T}_{ab}^i$  do not
have exactly the same commutation relations as those of the
dWHN-BFSS model.  Rather
we have $[\tilde{S}_{ab}^{i},\tilde{S}_{bc}^{i}]\neq 0$, as can
be easily checked explicitly.  
Thus, it is necessary to choose an operator ordering when
realizing (\ref{hatU}) as an operator.  We make the simplest
choice of symmetric ordering, which is shown in Fig. (\ref{orderingfig}). 
\begin{figure}
	\centerline{\mbox{\epsfig{file=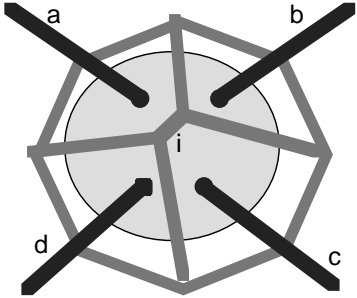,height=2in}}}
	\caption{The action of  the potential energy term,
	$Tr\left ( [\tilde{S}^{i},\tilde{S}^{j}]
[\tilde{S}_{i},\tilde{S}_{j}] \right )$ of the $D=3$ matrix model on
the boundary.  The action of the operator is to add the
shaded figure. In this and the following figures, 
unmarked wide shaded lines are in the adjoint representation.}
	\label{[][]fig}
\end{figure}

\begin{figure}
	\centerline{\mbox{\epsfig{file=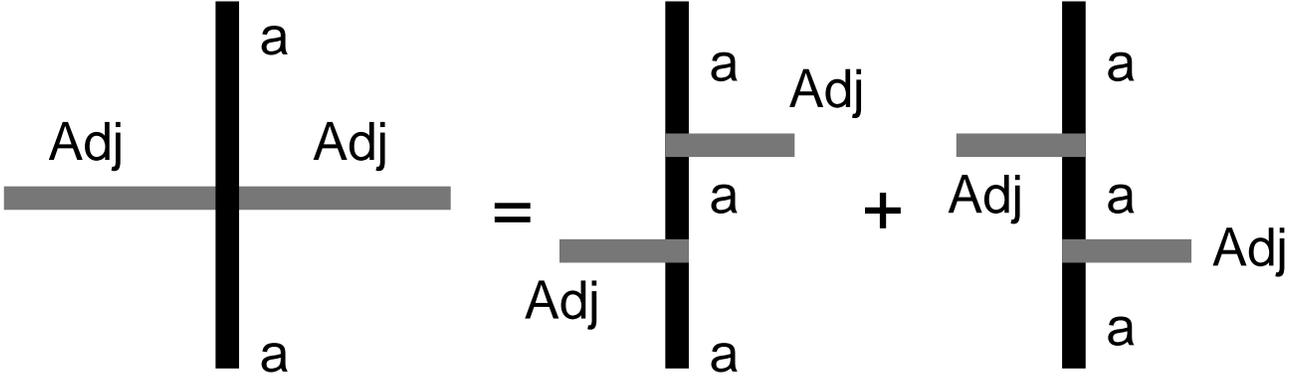,height=2in}}}
	\caption{The definition of symmetric ordering used
	in the definition of the potential energy
	operator shown in Fig. (\ref{[][]fig}).}
	\label{orderingfig}
\end{figure}
We can then form the potential energy operator (\ref{hatU})
by acting four times as indicated. The result is indicated
in Figure (\ref{[][]fig}), it is an operator that sums over all
$4$-tuples of edges $a,b,c,d$ and adds adjacent to the
punctures the graph indicated there.  The figure added may be
visualized as an $H$, with the external edges labeled by $Adj$'s
going to the four edges immediately adjacent to the nodes
$a,b,c,d$ and the cross-piece labeled by an $i\in Adj \times Adj$,
which is summed over.  The latter is the result of the
antisymmetrization in (\ref{hatU}). Finally, the overall weight is
\f
{1 \over g^{2}}w(i)
\ff
where in standard notation \cite{KL}
\f
w(i)= { \Delta_{i} \over \Theta(2,i,2) } \left (1-\lambda^{22}_{i}
\right )
\ff

\begin{figure}
	\centerline{\mbox{\epsfig{file=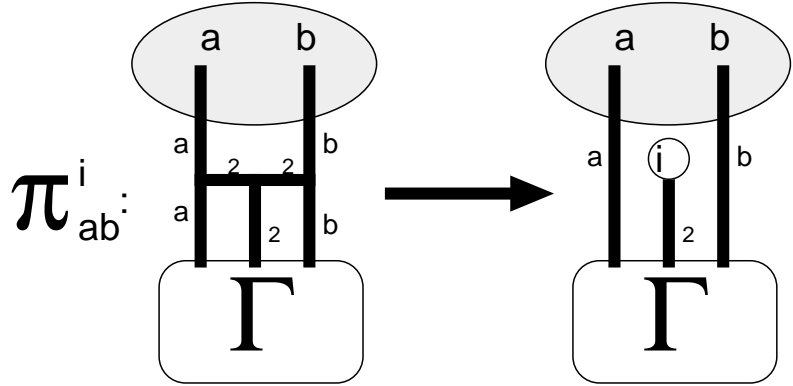,height=2in}}}
	\caption{The action of the momentum operator
	$\tilde{\Pi}_{i}^{ab}$. $2$ denotes the adjoint representation.}
	\label{piabifig}
\end{figure}

The next step is to invent the momentum operators
$\tilde{\Pi}_{i}^{ab}$ which are to be conjugate to the
$\tilde{S}_{ab}^{i}$.   We chose them so that, on
${\cal V}_{j_a}^{S^{2}}$
\f
[ \tilde{S}_{ab}^{i} , \tilde{\Pi}_{j}^{cd} ] = i\hbar 
\delta^{i}_{j} \delta^{c}_{a}\delta^{b}_{d}
\label{www}
\ff
$\tilde{\Pi}_{j}^{cd}$ can then be defined on a basis of
states in ${\cal V}_{j_a}^{S^{2}}$ given by intertwiners
of $G_{q}$, decomposed so that there are two framed edge of spin $r$
and $s$ 
linking the edges that go to $c$ and $d$  to a trinion,
which has a third edge of spin $u$ going 
outward(See Figure (\ref{piabifig}).)  
For every $c$ and $d$ such a basis can be constructed.
The
action of $\tilde{\Pi}_{j}^{cd}$ on this state is to remove 
the component where $r=s=u=Adj$ and replace it with
the state in which the now opened
edge $u$ is traced with $\gamma^{i}$.   
\begin{figure}
	\centerline{\mbox{\epsfig{file=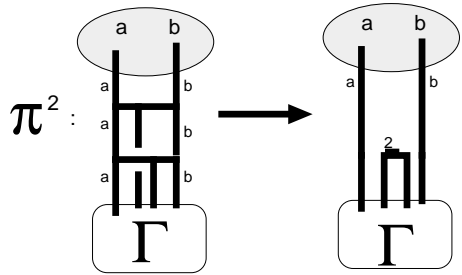,height=2in}}}
	\caption{The action of the kinetic  energy term,
	$Tr\left ( \tilde{\Pi}_{i} \tilde{\Pi}_{i} \right )$ 
	of the $D=3$ matrix model on the boundary.}
	\label{pi2fig}
\end{figure}
We then compute  the square of $\tilde{\Pi}_{i}^{ab}$.  The
action 
$\sum_{i}\tilde{\Pi}_{i}^{ab}\tilde{\Pi}_{i}^{ab}$
in the basis described is indicated in Figure (\ref{pi2fig}).

By adding the kinetic and potential energy together with the
weights given by (\ref{qH}), 
we  have the matrix Hamiltonian for the evolution of the
state in ${\cal V}_{j_a}^{S^{2}}$,
\f
\hat{H}= {g^{2} k^{4}\over 32\pi^{2}}Tr\left ( \tilde{\Pi}_{i} 
\tilde{\Pi}_{i} \right ) 
+ {1 \over g^{2}}
\left ( {2\pi \over k}  \right )^{8}
Tr\left ( [\tilde{S}^{i},\tilde{S}^{j}]
[\tilde{S}_{i},\tilde{S}_{j}] \right )
\label{boundaryH}
\ff
We can then form the evolution operator on 
${\cal V}_{j_a}^{S^{2}}$ as
\f
\hat{U}(s,t) = e^{i(t-s) \hat{H}}.
\label{boundaryU}
\ff

With this hamiltonian in hand, 
we are prepared to ask 
whether there is a choice of evolution move and amplitude
that reproduces its effect in appropriate circumstances.  
Let us begin with the simplest possible circumstance, which
is where the holographic surface bounds a small region,
containing only the parts of the surfaces ${\cal S}^{g}$
on which the replacement operator acts. In this
case the map $\Phi$ is trivial, and we can in the
right limits, establish an exact correspondence
between the replacement moves and their amplitudes
and the hamiltonian (\ref{boundaryH}). 

It is harder to
establish the correspondence for evolution
moves that act far from the holographic surface, and this
has not yet been done.  It is likely that this will be 
where supersymmetry plays a crucial role, to enforce 
a non-perturbative version of a non-renormalization theorem,
that will imply that one can move the evolution move from
close to the boundary to deep inside the bulk without
changing the form of the Hamiltonian that describes
the dynamics induced on the punctures in the boundary.  

For the present we restrict ourselves to the simplest
case, which is where the bulk contains the simplest possible
states on which the different terms of the matrix model
hamiltonian may act.  Because of the existence
of the free coupling constant $g$, the case where the
evolution move acts just inside the holographic surface
is sufficient to pick out a set of evolution moves
that match for both large and small $g$.

We are looking for a replacement move ${\cal R}_{matrix}$
of the form discussed in section 5 and an operator
$\hat{h}$ in the space ${\cal V}_{j_a}^{S^{2}}$ that
bounds the region where the replacement occurs.
In the simplest case, we may identify that surface with
the holographic surface on which the hamiltonian
(\ref{boundaryH}) acts.

We will consider first the limit $g^{2}\rightarrow 0$
and then the limit $g^{2}\rightarrow \infty$. 

\begin{figure}
	\centerline{\mbox{\epsfig{file=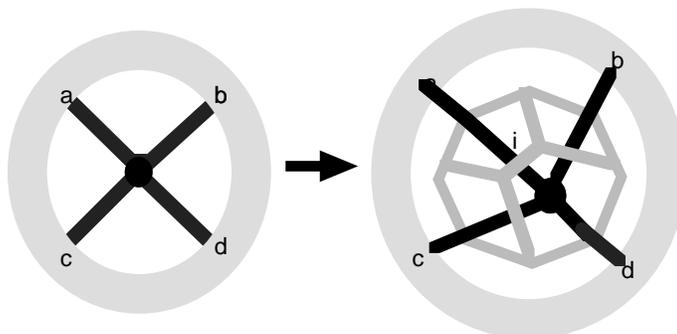,height=2in}}}
	\caption{The replacement move, $R_V$, in the bulk that
	induces the potential energy term,
	$Tr\left ( [\tilde{S}^{i},\tilde{S}^{j}]
[\tilde{S}_{i},\tilde{S}_{j}] \right )$ of the matrix model on
the boundary in the simplest case.}
	\label{[][]bulkfig}
\end{figure}

In the first case we are interested to match the potential
energy in (\ref{boundaryH}) to a single action of a replacement move.
The simplest
case to study is that where  $N=4$. We take the initial state in 
${\cal H}^{4}_{j_a}$ to be the simplest possible one,
indicated in Figure (\ref{[][]fig}) in which the four edges are tied together
by a four punctured sphere.  This state is labeled
by an intertwiner on the four punctured sphere, which we may
indicate simply as $|\mu>$. In this case the map 
(\ref{holomap}) commutes with the action of $\hat{H}$ because
the region cut out by the evolution move is the whole
interior.  The effect of the
evolution move may be arrived at by simply lifting the effect of
$\hat{H}$ from ${\cal V}_{j_a}^{S^{2}}$ to ${\cal H}^{N}_{j_a}$.
The kinetic energy term does not act, while the potential
energy term (\ref{hatU}) increases the genus by $\Delta g =8$, as shown
in Figure (\ref{[][]bulkfig}).  If the final state after the replacement move
is indicated by  $|Final >$ then in this case the amplitude is
given by 
\f
{\cal A}[\mbox{Figure (\ref{[][]bulkfig}) }] 
= <\tilde{Final} |e^{ \imath \hat{H}} |\tilde{\mu} >
\approx {i \over g^{2}}
\left ( {2\pi \over k}  \right )^{8} <\tilde{Final} |\hat{V} |\tilde{\mu} >
\ff
Here we have taken the time interval to be one because the initial
state has been completely replaced in the interior, corresponding
to one time step in any slicing.  

\begin{figure}
	\centerline{\mbox{\epsfig{file=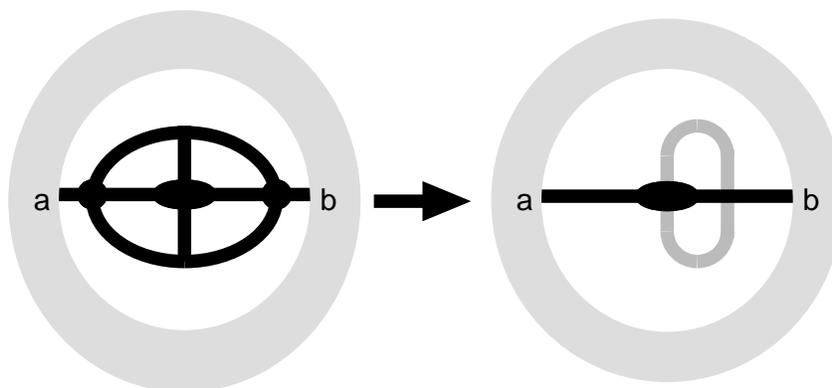,height=2in}}}
	\caption{The replacement move, $R_K$, in the bulk that
	induces the kinetic  energy term,
	$Tr\left ( \tilde{\Pi}_{i} \tilde{\Pi}_{i} \right )$ 
	in the simplest case in which it acts.}
	\label{bulkpi2fig}
\end{figure}

To see the effects of the kinetic energy dominate we consider
instead the limit where $g^{2}$ is large, in which case the leading
order term is its action on states in
${\cal H}^{N}_{j_a}$ where the interior surface has genus
$4$.  The action is shown in Figure (\ref{bulkpi2fig}).  The
change in genus is $\Delta g=-3$.   Indicating the
initial and final states in ${\cal H}^{4}_{j_a}$ by $|in>$ and $|out>$
the amplitude is 
\f
{\cal A}[\mbox{Figure (\ref{bulkpi2fig})}] = <\tilde{future} 
|e^{i \hat{H}} |\tilde{past} >
\approx {ig^{2} k^{4}\over 32\pi^{2}} <\tilde{future} | 
Tr \left ( \tilde{\Pi}_{i}^{ab}\tilde{\Pi}_{i}^{ab}
\right ) |\tilde{past} >
\ff

The replacement operator $\cal R$ must reduce to these two
limits.  The simplest choice that does this is to
take the sum of two operators 
\f
{\cal R}={\cal R}_{K} + {\cal R}_{V}
\label{Rsum}
\ff
To summarize, the 
first of these moves is a genus $\Delta g = -3$ move shown in 
Figure (\ref{bulkpi2fig}).
The amplitude for it is given by 
\f
{\cal A}[{\cal R}_{K}] = {ig^{2}k^{4} \over 32\pi^{2}} <\tilde{future} | 
Tr \left ( \tilde{\Pi}_{i}^{ab}\tilde{\Pi}_{i}^{ab}\right ) |\tilde{past} >
\label{amp1}
\ff
The second move is the genus $\Delta g=+8$ indicated in Figure 
(\ref{[][]bulkfig}),
to which we assign an amplitude 
\f
{\cal A}[{\cal R}_{V}] =  {i \over g^{2}}
\left ( {2\pi \over k}  \right )^{8}
<\tilde{future} |\hat{V}  |\tilde{past} >
\label{amp2}
\ff

These choices complete the definition of the theory for the
case $G_{q}=SU(2)_{q}$.  

\section{Extension to $Osp(1|2)$}

We now extend the matrix model to include fermions, keeping
for this section still to the $D=3$ case. This can be done by
extending the algebra to 
$G_{q}=Osp(1|2)$.  It is easy to see that the two moves
${\cal R}_{1}$ and ${\cal R}_{2}$ we have just defined
correspond to the bosonic part of the $D=3$ matrix model Hamiltonian.
The only adjustment that needs to be made is that the
spin $1$ and spin $1/2$ edges are replaced, respectively,
by the adjoint and fundamental representations of
$Osp(1|2)$.

We must then add a third replacement move to correspond to the
fermionic term in the matrix model. As we did with the
bosonic part we first construct the hamiltonian on the
boundary state space, ${\cal V}_{j_a}^{S^{2}}$, and then
pick an evolution move and amplitude that reproduces it.

The algebra $Osp(1|2)$ has five generators, which
are the three angular momenta, $J^{ij}$ and the supersymmetry 
generators $\hat{Q}^{A}$, where
$A=1,2$.  The algebra includes,
\f
[J^{ij} , Q^{A} ] = \sigma^{ij \ A}_{\ \ B} Q^{B}
\ff
where $J^{ij}$ are the generators of $Spin(3)$ and
$\sigma^{ij \ A}_{B}$ are its usual representation via
Pauli matrices.
The fundamental representation of $Osp(1|2)$ is three dimensional,
and comprises the spin $0$ and $1/2$ representations of
$SU(2)$.   We write its components as 
$\bar{A} = (0,1,2)= (0,A)$.  The adjoint representation is generated
by the graded symmetric generators $J^{(\bar{A}\bar{B})}$, where
\f
J^{(\bar{A}\bar{B})}={1\over 2} \left [ J^{\bar{A}\bar{B}} +
(-1)^{g(\bar{A})g(\bar{B})}J^{\bar{B}\bar{A}}
\right ]
\ff
where $g(0)=1, g(A)=0$.
The embedding of the generators in the fundamental representation is 
\f
\Gamma^{\bar{A}\bar{B}}= \left (   
\begin{array}{cc}
	\sigma^{AB} & Q^{B}  \\
	Q^{A} & 0
\end{array}
\right )
\ff

Thus,  $\sigma^{A}_{B}$ is imbedded in  the  $2\times 2$ 
block $\bar{A},\bar{B}=A,B$ of $\Gamma^{\bar{A}\bar{B}}$
and  
$[Q^{A}]_{\bar{A}}^{\bar{B}}$ is represented by 
$[Q^{A}]_{B}^{0}=[Q^{A}]_{0}^{B}= \delta_{B}^{A}$ and
$[Q^{A}]_{0}^{0}=0$.  

We may construct the
superconnection one-form
\f
{\cal A}^{\bar{A}}_{\bar{B}}= A^{AB} \sigma^{AB \ \ \bar{B}}_{\ \bar{A}} +
\lambda^{D}Q_{D \ \ \bar{B}}^{\ \bar{A}}.
\label{superconnection}
\ff
The phase space of the boundary theory is then given by the flat
superconnections on the punctured $S^{2}$ representing the boundary.
We then define the supersymmetric extensions of the $D0$ brane
coordinates on the phase space of the boundary theory to be
\f
\hat{T}^{AB}_{ab}= Tr\left ( \sigma^{AB}e^{\int_{\gamma_{ab}} {\cal A}}\right )
\label{S1}
\ff
\f
\hat{T}^{A}_{ab}= Tr\left ( Q^{A}e^{\int_{\gamma_{ab}} {\cal A} }\right )
\label{S2}
\ff

Quantum operators corresponding to these may be constructed
as we did in the last section,  using instead of $Spin(3)$
the fusion algebra of $Osp(1|2)_{q}$.  

The Hamiltonian is constructed from  two sets of operators, 
corresponding to the fundamental and the adjoint representation,
respectively.  To see how these correspond to the usual coordinates
of the matrix model we write the adjoint rep. operators as
\f
S^{\bar{A}\bar{B}}_{ab}= S^{\bar{B}\bar{A}}_{ab} = \left (   
\begin{array}{cc}
	S^{AB}_{ab} & \lambda^{B}_{ab}  \\
	\lambda^{A}_{ab} & 0
\end{array}
\right )
\ff
and the fundamental representation fields as
\f
\Psi^{\bar{A}}_{ab} = \left (
\begin{array}{c}
	\psi^{A}_{ab} \\
	\phi_{ab}
\end{array} \right )
\ff
We see that we have auxillary fields $\phi_{ab}$ and 
$\lambda^{A}_{ab}$ which are, in terms of $SU(2)$ reperesentations,
a matrix-valued scalar and spinor.

Conjugate to $S^{\bar{A}\bar{B}}_{ab}$ we have canonical momenta
$\Pi^{\bar{A}\bar{B}}_{ab}$, satisfying
\f
\{ S^{\bar{A}\bar{B}}_{ab}, \Pi_{\bar{C}\bar{D}}^{cd} \} = 
\delta^{(\bar{A}\bar{B})}_{\bar{C}\bar{D}}\delta^{c}_{a}\delta^{d}_{b}
\ff 
while the fermions satisfy
\f
\{  \Psi^{\bar{A}}_{ab}, \Psi_{\bar{B}}^{cd} \}_{+} = 
\delta^{\bar{A}}_{\bar{B}}\delta^{c}_{a}\delta^{d}_{b}
\ff
In terms of these $Osp(1|2)$ representations the supersymmetric
$D=3$ matrix model is
\f
H^{susy,D=3} = { g^{2 } \over 2} Tr \left [ 
\Pi^{\bar{A}\bar{B}}\Pi_{\bar{A}\bar{B}} \right ]
+ {1 \over g^{2}} Tr \left ( [S^{\bar{A}\bar{B}},S^{\bar{C}\bar{D}} ] 
[S_{\bar{A}\bar{B}},S_{\bar{C}\bar{D}} ] \right ) + 
Tr \left (\bar{\Psi}_{\bar{A}} 
\left [ \tilde{S}^{\bar{A}\bar{B}},{\Psi}_{\bar{B}}
\right ] \right ) 
\label{susy3H}
\ff
If one expands it is not difficult to see that the auxilary fields
$\phi_{ab}$ and $\lambda^{A}_{ab}$
decouple, leading to the $N=1$,  $D=3$ matrix model,
\f
H^{susy,D=3} = { g^{2 } \over 2} Tr \left [ 
\Pi^{i}\Pi_{i} \right ]
+ {1 \over g^{2}} Tr \left ( [S^{i},S^{j} ] 
[S_{i},S_{j} ] \right ) + 
Tr \left (\bar{\Psi}_{A}\sigma_{i}^{AB}
\left [ \tilde{S}^{i}, \tilde{\Psi}_{B} 
\right ] \right ) 
\label{susy3Hb}
\ff

\begin{figure}
	\centerline{\mbox{\epsfig{file=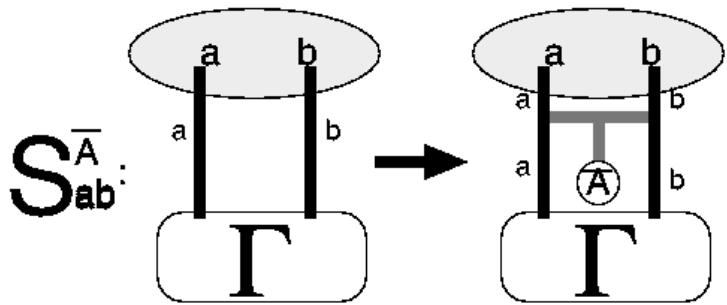,height=2in}}}
	\caption{The action of the fermionic operator corresponding
	to the fundamental representation of $Osp(1|2)$.  Note that the
	intertwinor shown is allowed in the representation theory of
	$Osp(1|2)$ as the condition that the spins must add up to
	be even is not satisfied\cite{superspin}.}
	\label{fsiab}
\end{figure}

We then proceed as before to represent these as operators 
in the space of intertwinors for the quantum deformation of
$Osp(1|2)$, which may be constructed as the space of states
of the $Osp(1|2)$ Chern-Simons theory corresponding to the
flat superconnections on the punctured surface.   
The operators that corresponds to $S^{\bar{A}\bar{B}}_{ab}$
and its momenta $\Psi^{\bar{A}}_{ab}$ are exactly those shown in
Figures (\ref{Sabifig}) and (\ref{piabifig}) 
where the adjoint representation of
$Osp(1|2)$ is now used.  
For the fermionic variables we use the operator shown in 
Figure (\ref{fsiab})

\begin{figure}
	\centerline{\mbox{\epsfig{file=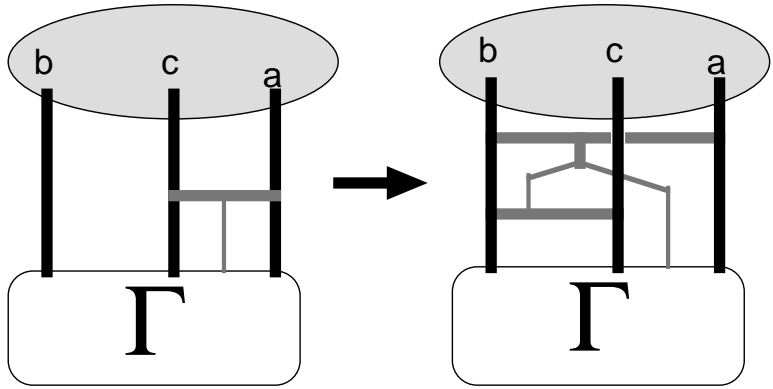,height=2in}}}
	\caption{One of two terms of 
	 the fermionic term of the matrix model,
	(\ref{Hf}).  This is the simplest state in which it
	can act. In the $D=3$ theory grey tubes are in the $3$-dimensional
	fundamental representation of $Osp(1|2)$ and thicker grey tubes are in the 
	$5$-dimensional adjoint representation. In the other term
	the order of the two thick grey tubes is reversed.}
	\label{ftermfig}
\end{figure}

The fermionic part of the matrix model hamiltonian is then expressed
as an operator on ${\cal V}_{j_a}^{S^{2}}$ by translating the 
expression
\f
H^{f}= { 4\pi^{2}\over k^{2}}Tr \left (\bar{\Psi}_{\bar{A}}
\left [ \tilde{S}^{\bar{A}\bar{B}}, \tilde{\Psi}_{\bar{B}} 
\right ] \right ) 
\label{Hf}
\ff
into the action shown in Figure (\ref{ftermfig})\footnote{Note that the
factor of $4\pi^{2}/k^{2}$ are determined as in the bosonic case
by the requirement that the operators that correspond to the
coordinates of the matrix model become commuting in the
limit $k \rightarrow \infty$.}.

\begin{figure}
	\centerline{\mbox{\epsfig{file=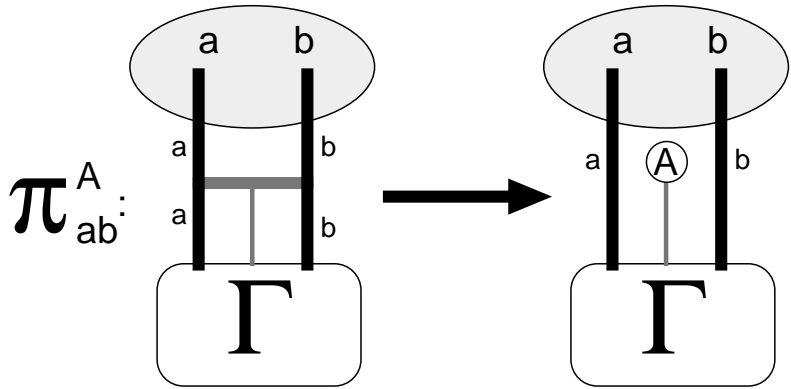,height=2in}}}
	\caption{The action of the fermionic momentum operator.}
	\label{fpifig}
\end{figure}

The fermionic momentum variable 
$\tilde{\Pi}_{\bar{A}}=\bar{\Psi}_{\bar{A}}$ is constructed 
by following the same procedure
we used for the bosonic momenta.  It yields the operator
shown in Figure (\ref{fpifig}).

\begin{figure}
	\centerline{\mbox{\epsfig{file=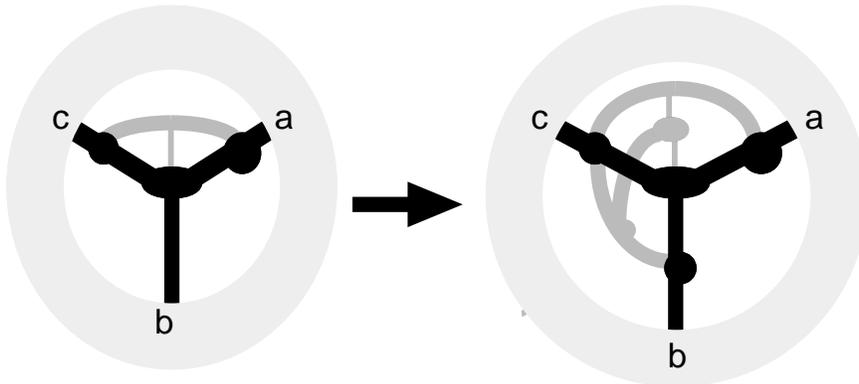,height=2in}}}
	\caption{The replacement move in the bulk that
	induces the fermionic term of the matrix model,
	(\ref{Hf}).}
	\label{Hfbulkfig}
\end{figure}

We then find a replacement operator which in the simplest
case reproduces (\ref{Hf}). The change of genus in this
case is $\Delta g = +1$.  It is given in Figure (\ref{Hfbulkfig}).
The corresponding amplitude is given by,
\f
{\cal A}[(\mbox{Figure }\ref{Hfbulkfig})] =  
i <\tilde{future} |H^{f}|\tilde{past} >
\label{famp}
\ff

The full replacement operator for $G_{q}= Osp(1|2)$
is then given by the $Osp(1|2)$ replacement operators
\f
{\cal R}= {\cal  R}_{K} + {\cal  R}_{V} + {\cal  R}_{f}
\label{freplace}
\ff

\section{The bosonic matrix model in $D=9$}

The procedure we have just used in $D=3$ can be used to construct
a background indepedent membrane theory associated with a matrix model
in any dimension $D$.  The basic idea in each case is that the
matrix coordinates are given by operators defined like the $S_{ab}^{i}$
in the space of intertwiners of the punctured sphere for some group
$G_{q}$. In the bosonic case $G_{q}$ will generally be the quantum
deformation of $SO(D)$ gotten by quantizing Chern-Simons theory on
a three manifold which is the punctured sphere cross the interval.
In the supersymmetric case the construction may be considerably
more complicated, as we saw in the $D=3$ case, where auxiliary
fields entered the construction.  In this case they decoupled but
this is not gauranteed to happen in the general case.
 
We now briefly sketch the $D=9$ case which is appropriate for the
dWHN-BFSS matrix model.  Full details will appear 
elsewhere\cite{mextended}.

\begin{figure}
	\centerline{\mbox{\epsfig{file=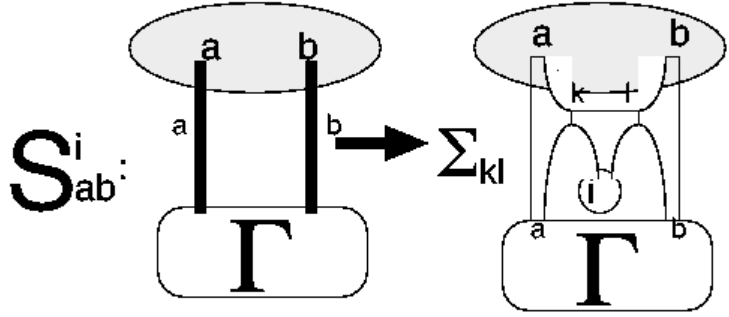,height=2in}}}
	\caption{The definition of the operator $\hat{S}^{i}_{ab}$
	in the $D=9$ case.
	The sums are 
	$k,l=V \oplus V^{2}\oplus V^{3} \oplus V^{4} = \Delta \otimes 
	\Delta -C $.  The $i$ stands for the projection of the vector
	representation $V$.  The remainder of the surface is
	indicated by $\Gamma$, other ends on the boundary are not
	shown.}
	\label{Sabi9fig}
\end{figure}

\begin{figure}
	\centerline{\mbox{\epsfig{file=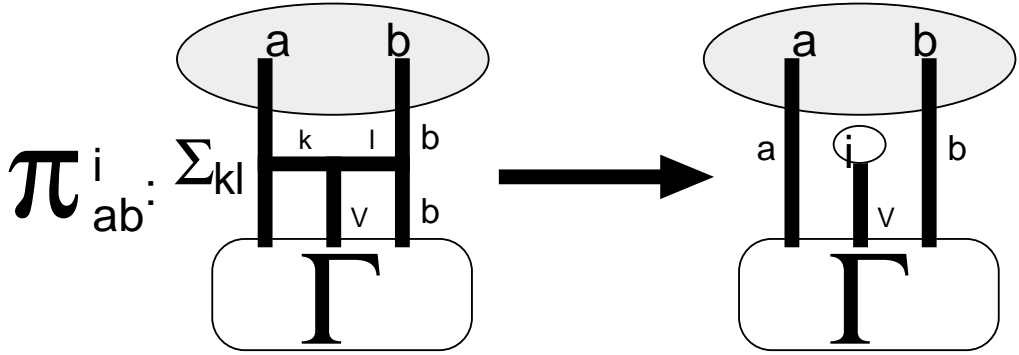,height=2in}}}
	\caption{The action of the momentum operator in the
	$D=9$ case.
	$\tilde{\Pi}_{i}^{ab}$.}
	\label{piabifig9}
\end{figure}

It is straightforward to extend the definition of the  $D=3$
bosonic matrix model to a bosonic matrix model in  $9$ dimensions,
by choosing for  $G_{q}$
the quantum deformation of  $SO_{q}(9)$.  The classical phase space
of the holographic observables will be given by the conformal blocks
constructed from the quantization of a $Spin(9)$ Chern-Simons theory.
The classical matrix model coordinates will be given by
(\ref{matrix17}) with the traces and the 
$\Gamma^{i}$ taken in the $16$ dimensional spinor representation
of $SO(9)$, which we denote by $\Delta$. The corresponding 
quantum operators $S^{i}_{ab}$ 
is shown in Fig. (\ref{Sabi9fig}), it is similar to the $D=3$ case,
but a bit  more
complicated.  The reason is that the  calculation of the quantum operator 
involves the decomposition of the product of spinor representations, 
and above $D=3$ this includes more than the adjoint and a scalar. In
$D=9$ we have,
$\Delta \otimes \Delta = C \oplus V \oplus V^{2}\oplus V^{3} \oplus V^{4}$
where $V=R^{9}$ and $V^{p}$ stands for the $p$-fold antisymmetric 
product.  With the trace part proportional to  $C$ removed, this 
replaces the adjoint representation when we extend from the 
$SU(2)$ case to the $Spin(9)$ case.
The momentum operator is defined in Fig. (\ref{piabifig9}), which 
works  just as in the $D=3$ case.

The potential energy and kinetic energy terms are modified 
appropriately, so that the adjoint representation of $SU(2)$
is everywhere replaced by the sum over the decomposition of 
$\Delta \otimes \Delta - C$.    The matrix model hamiltonian
is then represented in ${\cal V}^{{\cal S}^{2}}_{j_{\alpha}}$
by the hamiltonian (\ref{boundaryH}) and the simplest operator in the
bulk which induces it is again given by the two replacement operators
${\cal R}_{K}$ and  ${\cal R}_{V}$ as in (\ref{Rsum}), where in each
case the adjoint representation is replaced by the tracefree
sum $\Delta \otimes \Delta -C$.

\section{The supersymmetric extension in $D=9$}

What remains to arrive at a theory that includes the
full dWHN-BFSS matrix model is then only to include the fermionic degrees of
freedom.  To do this one must pick an appropriate superalgebra $G_{q}$ so that
one can write a superconnection of the form 
(\ref{superconnection}) corresponding to the
supersymmetric extension of the matrix model.  The
spinor representation of $SO(9)$ is $16$ dimensional, to incorporate
one spinor degree of freedom what is required is a superalgebra with
a $17$ dimensional fundamental representation whose bosonic subalgebra 
has a $Spin(9)$ subalgebra.  

There are several choices  for $G_{q}$ which each give rise to a theory
whose degrees of freedom include those of the dWHN-BFSS matrix
model. One of the simplest, which involves only one fermionic matrix
variable, is
$SU(16|1)$.  Given that the norm in the spinor representation of
$Spin(9)$ is symmetric we have the inclusions, 
$Spin(9) \subset SO(16) \subset SU(16)$.  This has a fundamental
$17$ dimensional representation, which is the smallest
supermultiplet of  $Spin(9)$ comprising the scalar and the spinor.
This is the analogue of the $3$ dimensional representation of
$Osp(1|2)$ used in the supersymmetric $D=3$ matrix model and in
supersymmetric spin networks in $D=3+1$\cite{superspin}.

A consequence of the unification in a larger group, however is the
inclusion of additional bosonic degrees of freedom. This is seen
by the fact that the symmetric product of spinor representations
of $Spin(9)$ involves
\f
16 \otimes_{sym} 16 = R \oplus V \oplus V^{4} = 1 \oplus 9 \oplus 126
\ff
The $126$ is the four index antisymmetric tensor in $D=9$.
In terms of indices this corresponds to
\f
S^{A}\otimes_{sym} S^{B} = \delta^{AB}\phi \oplus \Gamma^{AB}_{i} V^{i}
\oplus \Gamma^{AB}_{ijkl} V^{ijkl}
\ff
where $ \Gamma^{AB}_{abcd}$ is the antisymmetric product of $9$
dimensional Gamma matrices $\Gamma^{AB}_{a}$.
This means that the matrix model degrees of freedom have been
extended from $X^{i}_{ab}$ to $X^{(AB)}_{ab}$, so as to include
naturally the four (or, by duality, five) index antisymmetric
tensor fields and a scalar:
\f
X^{(AB)}_{ab}= \delta^{AB}S_{ab} + \Gamma^{AB}_{i} X^{i}_{ab}
+ \Gamma^{AB}_{ijkl} V^{ijkl}_{ab}
\ff
The resulting theory involves then the addition of central
charges to the $D=9$  superalgebra,
\f
\{ Q^{A} , Q^{B} \} = \delta^{AB}  +H  \Gamma^{AB}_{i} P^{i}
+ \Gamma^{AB}_{ijkl} Z^{ijkl} 
\ff
This is, of course, not surprising as the usual matrix model 
superalgebra closes on shell, and we are working in a formalism in 
which the supersymmetry should be manifest.  It is possible
that these additional degrees of freedom provide an alternative
way for the transverse five-brane to be represented in the matrix
model; this will be discussed elsewhere\cite{mextended}.

One can then define the theory in terms of the same three
replacement moves 
as the $Osp(1|2)$ case, indicated in (\ref{freplace}), 
with the appropriate replacements of $Osp(1|2)$ representations with
$SU(16|1)$ representations.
The adjoint representation of
$Osp(1|2)$ is replaced the tracefree symmetric tenor representation of $SU(16|1)$
(which contains the vector and $4$-form of $SO(9)$)
and the $3$-dimensional fundamental representation of
$Osp(1|2)$ is replaced by the $17$-dimensional fundmamental
representation of $SU(16|1)$. 

An alternative supersymmetric extension of the bosonic
$9$ dimensional theory arises from choosing $G_{q}= Osp(16|2)$.
This has a bosonic subalgebra $SO(16) \oplus Sp(2)$. This doubles
the supersymmetry to an $SU(2)=Sp(2)$ doublet of  supersymmetry
charges $Q^{A}_{I}$, $I=1,2$.  

In either case it is then natural to consider extending the theory
by choosing for $G_{q}$ the algebra $Osp(1|32)$, which has been suggested 
is the full supersymmetry algebra of $\cal M$ theory\cite{Osp32}.
It is possible that this gives rise to a covariant formulation 
of the matrix model, this will also be discussed in \cite{mextended}.

\section{Conclusions}

We conclude by summarizing the main results and conjectures, after 
which we indicate some directions for future work.

\subsection{Summary of results}

W began this paper with a summary of the construction of a class of background
independent quantum theories of gravity described in 
\cite{tubes,pqtubes}. These theories are found by extending the 
spaces of states and histories of quantum general relativity and
supergravity so that the algebra of observables are based on an
arbitrary quantum group, $G_{q}$.  The  
space of states is constructed from the spaces of conformal
blocks on all finite genus two-surfaces, which gives rise to
their interpretation as background independent membrane field theories.
The dynamics is expressed in the causal histories framework\cite{FM3}
which describes dynamics in terms of a sum over histories, each
of which is endowed with a discrete causal structure.  

The main results of this paper concern a class of observables, which
are defined on boundaries.  The boundary state spaces are spaces
of conformal blocks, or intertwiners, on punctured surfaces.  
As these observables are known in $3+1$ quantum general relativity and
supergravity, in the presence of certain boundary conditions,
to coordinatize a certain, perhaps complete, subspace of physical states, 
we call them holographic observables.   In these cases these boundary
state spaces have dimensions which grow exponentially as the area of
the boundary, as required by the Bekenstein bound. We
found that a complete set of observables on these boundaries may
be used to construct a generalization of the matrix models, in which
there are non-trivial commutators, proportional to $1/k$ between
the matrix coordinates. We argued that the cosmological constant
vanishes in the limit $k \rightarrow \infty$, as has been shown quantum general
relativity and supergravity in $2+1$ and $3+1$ dimensions.   
This led to the suggestion that the standard matrix models may
be reproduced in the limit $k \rightarrow \infty$.  
For  $G_{q}=Spin(D)$ the holographic observables
$\hat{S}^{i}_{ab}$ provide a representation of a matrix-like
system in the boundary state space  ${\cal V}^{{\cal S}^{2}}_{j_{\alpha}}$
that in the limit $k \rightarrow \infty$ and large 
$N$ will reproduce the degrees of freedom and dynamics of the matrix
models (\ref{classH}).  We have seen that in $D=3$ the choice
$G_{q}=Osp(1|2)$ leads naturally to its supersymmetric extension
and that in $D=9$ the choice $G_{q}=SU(16|1)$leads to a theory which 
includes the dWHN-BFSS matrix, model, with some additional degrees of
freedom.  

There were additional results concerning the relationship between
the bulk theory and the boundary theory.  The dynamics of the bulk 
theory is expressed in the causal histories framework in terms of
a set of local moves, to which the theory assigns amplitudes.  We 
posited a particular bulk to boundary map between the infinite 
dimensional space of bulk states and the finite dimensional space of
boundary states.  In quantum general relativity and supergravity in
$3+1$ this map is known to express the hamiltonian constraint, we
posited here that it holds generally.  Given this map the question 
arises as to whether the dynamics of the  bulk theory can be chosen so 
that  the matrix dynamics is induced for the boundary observables.
We showed that this is true in both the bosonic and supersymmetric 
cases we studied in $D=3$ and $D=9$.

These were the main results of this paper.  We made in addition
several conjectures concerning the relationsip of these
results to $\cal M$ theory.

\subsection{Conjectures}

In sections 3-5 we argued that the holographic 
observables defined on ${\cal V}^{{\cal S}^{2}}_{j_{\alpha}}$ may be
interpreted as holographic observables in the sense of \cite{screens}.
This means that they receive information about the information on the
maximal past slicing.  In the continuum limit, if there is one, this
maximal past slicing becomes the past lightcone of the boundary $\cal 
S$ on which the holographic observables are defined.

The key problem in any background independent approach to quantum 
gravity is to show that there is a continuum limit which reproduces
the semiclassical description of general relativity coupled to some
set of quantum fields or strings.  While there are promising results
so far in $1+1$ dimensions\cite{janrenata} we do not yet have the tools
to show this in the present case.  It must then be our first 
conjecture:

\begin{itemize}

	\item{}{\bf C1} The background independent membrane theory described above
	with $G_{q}$ taken to be the quantum deformation of $SU(16|1)$
	and dynamics given by the three replacement moves
	${\cal  R}_{K} + {\cal  R}_{V} + {\cal  R}_{f}$ has a set of
	continuum limits, corresponding to consistent backgrounds
	of $\cal M$ theory including  flat $11$ dimensional spacetime.
	
\end{itemize}	

In this case, the argument in section $3$ then tells us that the 
holographic surfaces ${\cal V}^{{\cal S}^{2}}_{j_{\alpha}}$ should represent 
a screen in flat $11$ dimensional spacetime, which recieves 
information from the past light cone.  Given that the matrix  
coordinates $\hat{S}^{i}_{ab}$ can be interpreted in the limit 
$k\rightarrow \infty$ and $N \rightarrow \infty$ to reproduce
the matrix model we may deduce the second conjecture.

\begin{itemize}

	\item{}{\bf C2} In that continuum limit, and in the limits
	$k,N \rightarrow \infty$, the matrix observables
	$\hat{S}^{i}_{ab}$ describe the embedding of the punctured
	two surface on which the holographic observables are defined in the
	$R^{9}$ which represents the transverse directions in the lightcone
	coordinates of an observer in $10+1$ spacetime.
	
\end{itemize}	

We have also made a third conjecture

\begin{itemize}

	\item{}{\bf C3} The theory with $G_{q}= SU(16|1)$ is a broken
	phase of a theory with $G_{q}= Osp(1|32)$ which will be a
	background independent form of $\cal M$ theory.  Its holographic
	observables will provide a covariant extension of the
	matrix model in $D=11$ and it will contain other phases 
	corresponding to the consistent perturbative string
	theories.
	
\end{itemize}
 
\subsection{Directions for future work}

Much needs to be done in order to develop the theory described
here so that the conjectures just made can be refuted or confirmed.
I close with a description of some of the questions that may be
attacked.

\subsection{The existence of the continuum limit}

The main dynamical problem yet to be attacked  is the existence of the
continuum limit.  A fundamental point is that, if there is a continuum 
limit, the causal structure induced on $\cal M$ will corresponds
directly 
to that of the actual spacetime geometry.  This makes the problem of
the continuum limit very different from that of a Euclidean theory.
One consequence is that 
the  continuum limit in such a
theory cannot be
described in terms of the conventional second order
critical phenomena\cite{fmls1,stulee}.
This is because the causal structure is a degree of freedom, so there
is no fixed metric to define a Euclidean continuation.
Instead it is analogous to non-equilibrium critical phenomena such as
directed percolation or the growth of soap bubbles\cite{fmls1,stulee}. 

Detailed studies 
such lorentzian critical phenomena for quantum theories of gravity
are in progress in $1+1$\cite{janrenata,sameer1} and $2+1$ 
dimensions\cite{roumensameer}.  One result is
the discovery of a universality class for $1+1$ quantum
gravity that is not Liouville theory\cite{janrenata}.  
Of great importance
would be the extension of these methods to the study of supersymmetric
theories. In this regard it is interesting to note that there exists
a possible connection\footnote{Pointed out by J. Ambjorn and 
S. Gupta.} between supersymmetry and causality in $1+1$ dimensional
matrix models of evolving causal histories.

These studies are necessary preliminaries for an attack on the
continuum limit of the full theory.  In particular, control
over the continuum limit of $1+1$ dimensional theories
of this kind may be sufficient to  discover those choices
of algebras $G_{q}$ and evolution laws for which the continuum
limit exists.  This is because, according to the argument of
\cite{stringsfrom}, summarized above, a causally evolving
spin network (or spin foam) cannot have a continuum limit
unless the $1+1$ dimensional spin system that describe its
small perturbations itself has a continuum limit which
reproduces a critical string theory.  

It is also 
intriguing that non-equilibrium critical phenomena of the kind
we expect to find in these theories, can be self-organized, in the
sense of Bak et al\cite{soc}, leading to the possibility of self-tuning
critical phenomena. It is intriguing to speculate that the self-tuning
may extend to the selection of algebras (or more properly,
subalgebras) such that continuum limits exist.  If this is the
case than the existence of the classical world may be a
self-organized critical phenomena.

One thing which is unusual about the continuum limit in this
theory is that it will have to pick out the dimension of space. 
At the quantum level generic states can be decomposed in 
different ways as superpositions of states that correspond
to pseudomanifolds of definite dimension.
It is interesting to speculate on the mechanisms whereby the
different bases which are dual to pseudomanifolds of different
dimensions may decohere in the case of very high genus, perhaps
leading to a  statistical theory of compactification.

\subsection{Extensions of the matrix model}

The results described above suggest that there ought to be
an extension of the dWHN-BFSS matrix model which incorporates
four form degrees of freedom and a central extension of the
supersymmetry algebra.  This is under investigation\cite{mextended}.
This may also lead to a supercovariant formulation of the matrix
model.

\subsection{Uniqueness of the evolution moves}

In the present paper we were interested just to show that 
there exist choices of evolution moves and amplitudes such that
the matrix model was recovered for the description of the
holographic degrees of freedom.  The form of the laws we
have found here is neither beautiful nor unique. It is then
interesting to investigate whether there are more elegant
forms for the evolution laws.

Of great importance is the question of whether the 
bulk moves still induce the matrix hamiltonian
when they act deep inside a very high genus surface.
One may conjecture that supersymmetry will protect
the form of the induced boundary action, due to
a non-perturbative renormalization theorem.  

\subsection{Defects and branes}

As mentioned in section 2, the correspondence between
quantum geometries in $D$ dimensions and states in
$\cal H$ reproduces pseudomanifolds.  These have
defects which are identified surfaces of codimension
$2$ or greater. Given the fundamental dynamics on
$\cal H$ these defects will evolve.  These may then be
something like the  branes of string theory.  
These are not $D$-branes
because the excitations which we have identified as strings
do not end on them, rather they must be magnetically
charged.  Thus, the theory predicts that there are $D0$
branes associated with the holographic surfaces on which
strings end, and dual magnetic branes, associated with
defects that arise in the dual transformation between
conformal blocks and combinatorial pseudomanifolds.

\subsection{Symmetry breaking and algebraic compactification}

There is a natural mechanism for symmetry breaking in this
class of theories, which is the following.  Let us consider
a basis state $|\Gamma , j_{\alpha}, \mu_{i} >$ for an
open tube network, with all the representations chosen
to be in the representation associated with $\wedge^{r} V$,
where $V$ is the fundamental spinor representation
of $Spin(D) \in G_{q}$.
Then small perturbations live in the $D-r$ dimensional subspace
of $V$. The result is that the algebra that describes the
small perturbations is reduced from $Spin(D)$ to $Spin(D-r)$.

The simplest example is in the case that $j$ is in the
fundamental representation, in which case as we saw above, small
perturbations live in the $D-1$ dimensional subspace associated
with the tangent space to $S^{D-1}$.  The symmetry group is
then reduced to $Spin(D-1)$.  For the case $D=3$, $Spin(3)=SU(2)$
and the small perturbations in the presence of $j=1/2$
are associated with $U(1)$.   In the general case
$G_{q}$ will be reduced to a subalgebra, $H_{q}$.

This means that the theory may have in principle many continuum
limits, associated with different reductions of $G_{q}$.  This
may lead to a theory of ``compactifications'' of the theory
that is entirely algebraic.

\subsection{Relationship to the AdS/CFT conjecture}

The basic picture which has emerged is that of an algebraic
description of a membrane whose dynamics, given the map $\Phi$,
may be described in a dual picture given by the  embedding of $N$
points in $S^{D-1}$.  This sphere is understood to be the
transverse surface associated with the past light cone of
some observer. In the limit $N \rightarrow \infty$ this
surface may be pictured itself as a two dimensional membrane
embedded in $S^{D-1}$.  In the presence of a symmetry
breaking state of the kind just defined, this will reduce
to the embedding of a membrane in $S^{D-r-1}$.  

We also note that the whole picture is well defined when the
algebra $G_{q}$ has been quantum deformed and $q$ is taken at
a root of unity. This is because in this case the list
of representations is finite, which means that the sum over
histories that defines the amplitudes of observable
processes (observable with respect to observers inside the universe)
is finite. Thus, the level $k$ defines a kind of infrared cutoff.

It is then not surprising that in $2+1$ and  $3+1$ dimensions, where
the correspondence between the quantum theories and the classical
theories are well understood, the level $k$ is associated with
the cosmological constant. In $3+1$ dimensions the exact
relation is given by (\ref{cc}) \cite{linking,hologr}.
We may also note that in this case the area of the 
$S^{2}$ grows proportionally to $N$.  As discussed in
section 2, we may choose the area operator in higher dimensions
so that this is true as well for the $S^{D-r-1}$ that the
membrane embeds in, in that limit.

Let us then consider the limit $N \rightarrow \infty$,
in which we keep the cosmological constant, and hence $k$,  fixed.  Let us
also assume that we have picked the sign of the 
cosmological constant, as well as the pattern of
symmetry breaking,   so that if there is
a continuum limit, it may be a $D-r+1$ dimensional
AdS spacetime.  It then follows that in the limit
$N\rightarrow \infty$ the $S^{D-r-1}$ in which the
holographic surface embeds must become the boundary
of a spatial slice of the AdS spacetime. This is because
its area will go to infinity and, by the symmetry of the classical
limit, it will be a metric sphere of that area.  The only
such embedded metric spheres with infinite area are
in the boundary.

If the AdS/CFT conjecture is also true then we reach the following
conclusion.  There is a inclusion, 
\f
{\cal F}: H_{AdS} \rightarrow {\cal V}_{\infty}^{S^{2}}
\ff
where $H_{AdS}$ is the physical Hilbert space of the conformal
field theory that describes the degrees of freedom at the boundary
of $D-r+1$ dimensional $AdS$ spacetime 
and ${\cal V}_{\infty}^{S^{2}}$ is the $N\rightarrow \infty$
limit of the space of conformal blocks ${\cal V}_{j_a}^{S^{2}}$
of the $N$ punctured $S^{2}$ for the reduced algebra $H_{q}$. 

In the $3+1$ dimensional case, this implies that the boundary
states for quantum general relativity and supergravity with a cosmological
constant should be constructible in terms of conformal blocks
on a punctured $S^{2}$, in the limit that the number of punctures
is taken to infinity.  This is confirmed in detail, for the reduction
down to general relativity and  ${\cal N}=1, 2$ 
supergravity\cite{linking,hologr,superholo}
For the full ${\cal N}=8$ case, the conjecture is that
the state space of the boundary theory 
lives in the $N\rightarrow  \infty$ limit of the space of
intertwiners of $Osp(2|4)\oplus Osp(2|4) $.  This
will be discussed in more detail in \cite{adsconj}.

\subsection{Black hole state counting}

An important problem that this framework should open up a solution
for is the entropy of black holes far from extremality, where
the specific heat is negative.  These are non-BPS states,
however, the present framework suggests that black hole
boundary states should still be expressible in terms of
$D0$ branes on an appropriate boundary.  Evidence for this
is given by the black hole state countings in \cite{kirill-bh}
which use a modification of the Chern-Simons boundary conditions
in \cite{linking,hologr}.  To study this one writes a path
integral formula for the thermal 
partition function of the theory.  
This will define the statistical mechanics of the $D0$ branes.  
This will be discussed in more detail elsewhere.

\subsection{Quantum cosmology from the inside}

The class of theories we have discussed
here are examples of a new kind of quantum theory of cosmology in which
there is no observables that acts directly on the 
``wavefunction of the universe''. This is good, because were there,
we would have to worry about who could observe them.  Instead, given
a quantum history $\cal M$, or a class of such histories, there
are a large number of Hilbert spaces that describe what observers
inside the universe may observe.  This is of course
closely related to the holographic principle, as discussed
here and in \cite{tubes,hic,screens}.  The general form of
such theories is discussed in \cite{FM3,FM4,FM5} and interpretational
issues  are discussed in \cite{cqc}. 

What is most gratifying about this class of theories
is the manner in which those structures which seem necessary
to solve foundational issues, such as how to describe what
observers inside the universe may measure in a quantum cosmology,
turn out to coincide precisely with those structures which 
are necessary to solve technical problems concerning the existence
of background independent quantum theories of gravity with
good continuum limits.

\section*{ACKNOWLEDGMENTS}

This work arose out of joint work with Fotini Markopoulou, whose
suggestions and criticisms have been crucial. The holographic
map $\Phi$ described here is a realization of ideas first
proposed by Louis Crane\cite{louis-holo} concerning 
the relationship between bulk
and boundary states in quantum gravity, and his picture has been
an important inspiration for the work leading up to the present
conjecture.
I am also indebted
to Arivand Asok, Chris Beetle, John Brodie, Surya Das, Michael Dine, 
Brian Greene, Sameer Gupta, Eli Hawkins, Chris Isham, 
Stuart Kauffman,  
Yi Ling, Shahn Majid, Seth Major, Juan Maldacena, 
Roger Penrose, Michael Reisenberger,
Carlo Rovelli, Steve Shenker and Edward Witten 
for conversations and criticisms at various
stages of this work.  I am especially indebted to Louis Kauffman
for pointing out an error in a construction in the first draft of
this paper. I am grateful to ITP in Santa Barbara and 
the Theoretical Physics Group at Imperial College for hospitality
during the course of this work.  
This work was supported by the NSF through grant
PHY95-14240. I am also very
grateful for support and encouragement from the Jesse 
Phillips Foundation.

\end{document}